\documentclass{article}

\usepackage{arxiv}

\usepackage[utf8]{inputenc} 
\usepackage[T1]{fontenc}    
\usepackage{hyperref}       
\usepackage{url}            
\usepackage{booktabs}       
\usepackage{amsfonts}       
\usepackage{amsmath}
\usepackage{amssymb}
\usepackage{nicefrac}       
\usepackage{microtype}      
\usepackage{lipsum}
\usepackage{graphicx}
\usepackage{dsfont} 

\usepackage{subcaption}
\usepackage{physics} 
\usepackage{multirow}
\usepackage{xcolor}
\usepackage{hyperref}

\usepackage{makecell}
\newcounter{magicrownumbers}
\newcommand\rownumber{\stepcounter{magicrownumbers}\arabic{magicrownumbers}}

\graphicspath{{./figure/}}

\title{Attractor Selection in Nonlinear Energy Harvesting Using Deep Reinforcement Learning}

\author{
	Xue-She Wang \& Brian P. Mann \\
	Dynamical Systems Research Laboratory \\
	Department of Mechanical Engineering \& Materials Science \\
	Duke University \\
	Durham, NC 27708, USA \\
	\texttt{xueshe.wang@duke.edu} \\
}

\begin{document}
\maketitle

\begin{abstract}
Recent research efforts demonstrate that the intentional use of nonlinearity enhances the capabilities of energy harvesting systems. One of the primary challenges that arise in nonlinear harvesters is that nonlinearities can often result in multiple attractors with both desirable and undesirable responses that may co-exist. This paper presents a nonlinear energy harvester which is based on translation-to-rotational magnetic transmission and exhibits coexisting attractors with different levels of electric power output. In addition, a control method using deep reinforcement learning was proposed to realize attractor switching between coexisting attractors with constrained actuation.
\end{abstract}


\section{Introduction}

For small and independent devices where replacing a battery or connecting to a power grid is unrealistic, such as environmental monitoring systems \cite{xi2017multifunctional,bastien2009ocean}, medical implants \cite{donelan2008biomechanical, paulo2010review} and structural health monitoring sensors \cite{park2008energy, galchev2011harvesting}, vibratory energy harvesters are useful to realize self-powering by converting mechanical energy to electrical energy with electromechanical coupling, e.g.~piezoelectric, electromagnetic, electrostatic transduction, etc. Traditional linear energy harvesters which operate based on linear resonance work well only when excitation frequency is close to the natural frequency. When the excitation frequency is known a priori, the geometry and dimensions of a linear harvester can be carefully selected to match its resonant frequency to the excitation frequency. However, when the excitation frequency is unknown or varies in different operational conditions, the harvester with pre-tuned resonant frequency is unable to achieve optimal power output \cite{tang2013broadband}. 

Many of the control methods applied to energy harvesters have primarily focused on resonant frequency tuning, which actively or passively matches the resonant frequency to environmental forcing frequency to increase harvesting efficiency. The active methods require continuous power input for resonance tuning. While for the passive methods, intermittent power is input for tuning and no power is required when frequency matching is completed, that is until the excitation frequency varies again \cite{roundy2005toward}. Resonance tuning methods can be categorized into mechanical, magnetic, and piezoelectric methods. Mechanical methods are usually developed based on elementary of vibration theory: the resonance of a system can be tuned by changing the stiffness or mass \cite{gu2010passive, jo2011passive}. Magnetic method is an option for resonance tuning by applying magnetic force to alter the effective stiffness of a harvester \cite{reissman2009piezoelectric, zhu2008closed}. Piezoelectric transducers provide another option for resonance tuning based on the fact that the stiffness of the piezoelectric material itself can be varied with various shunt electrical load \cite{peters2009closed, lallart2010frequency}.

Although the aforementioned control systems were implemented for tuning energy harvesters, they were not completely self-powered. The power required to apply control actions is oftentimes greater than the power harvested, thus an external power supply is still needed and they are only suitable for vibration scenarios where the forcing frequency changes infrequently \cite{mann2019intentional}. An alternate solution to achieving broadband frequency response is using nonlinear energy harvesters. Several recent works have suggested the intentional use of nonlinearity might be beneficial to energy harvesting systems~\cite{mann2009energy, mann2010investigations, erturk2009piezomagnetoelastic, stanton2009reversible}. These studies have explored the use of nonlinearities to broaden the frequency spectrum, extend the bandwidth, engage nonlinear resonances, and/or to facilitate tuning~\cite{renno2009optimal, triplett2009effect, wu2014energy, stanton2012melnikov, stanton2010nonlinear, ramlan2010potential, bowers2009spherical}. 

While these investigations, along with many other recent works, have advanced the current understanding on the beneficial use of nonlinearity, the introduction of nonlinearity can also cause many additional difficulties. Paramount amongst these challenges, and a common issue in nearly all nonlinear harvesting systems, is the presence of coexisting solutions~\cite{wang2020constrained}. More specifically, for a certain environmental excitation, there exists multiple responses of a harvesting system and thus  various levels of power generation~\cite{mann2019intentional}. These responses are stable steady-state oscillations, thus also considered ``attractors'' in nonlinear dynamics. An energy harvester generally prefers running on a higher-power attractor for faster energy harvesting, but would also need a lower-power attractor for safety reasons or physical restrictions. When one of the attractors is desirable and the other undesirable, it becomes critically important to have a control method to select the desired attractor with minimal energy expenditure.


Unlike conventional linear control problems where the system model is mathematically well-described or weakly nonlinear control problems (e.g. Lyapunov's function can be easily found), the control method of switching attractors should push the system far away from an equilibrium thus a highly nonlinear behavior is inevitable. Conventional control techniques, such as proportional-integral-derivative (PID) control and linear quadratic regulator (LQR) control, may not provide a robust control. In addition, the attractor switching for an energy harvester needs 1)~to apply constraints on control input, and 2)~an optimal control method to minimize the energy consumed during the control process. For example, the limitations on the instantaneous power/force and total energy/impulse of a controller need be considered in practice. Another practical consideration is the optimization of total time and energy spent on the control process. Switching attractors using as little energy or time as possible is oftentimes required, especially when attempting to escape detrimental responses or in the case of a finite energy supply.

Fortunately, a technique that is compatible with a broader scope of nonlinear systems, reinforcement learning (RL), can be applied without the aforementioned restrictions. By learning action-decisions while optimizing the long-term consequences of actions, RL can be viewed as an approach to optimal control of nonlinear systems~\cite{Chen1996RL}. Various control constraints can also be applied by carefully defining a reward function in RL~\cite{Sutton1992RL}. In recent years, a large number of advanced RL algorithms have been created to address complex control tasks, including benchmark works using MuJoCo physics simulator~\cite{Todorov2012Mujoco, Levine2013Guided, Schulman2015Trust, Heess2015Learning, Schulman2015High}, motion planing of robotics~\cite{Mahmood2018Benchmarking}, autonomous driving~\cite{Lange2012Autonomous}, and active damping~\cite{Turner2020RL}. Especially for the investigation of attractors in nonlinear dynamical systems, Ijspeert et al.\ proposed a generic modeling approach for attractor behaviors of autonomous nonlinear dynamical systems with the help of statistical learning techniques~\cite{ijspeert2013dynamical}. Wang et al.\ realized switching between coexisting attractors in a hardening Duffing oscillator using two RL algorithms: the cross-entropy method (CEM) and deep deterministic policy gradient (DDPG)~\cite{wang2020constrained}. Several researchers also have explored RL-based optimal control for gene regulatory networks (GRNs), with the goal of driving gene expression towards a desirable attractor while using a minimum number of interventions~\cite{Datta2003External, Sirin2013Employing, Imani2017Control, Papagiannis2019Deep}. 

The paper proposes a novel nonlinear energy harvesting system and a RL-based control method to switch between the attractors with different energy output levels. The content of this paper is organized as follows. Section~\ref{sec:system_design} presents the mathematical model of an energy harvester which is based on a translational-to-rotational magnetic transmission and electro-magnetic coupling. In Section~\ref{sec:attractors}, two types of coexisting attractors were observed with different levels of electric power output: high-power attractor and low-power attractor. Section~\ref{sec:rl_framework} presents the reinforcement learning framework designed for attractor switching of an energy harvester. Several key terms in RL were defined, including ``environment'', ``action'', ``state \& observation'' and ``reward''. Sections~\ref{sec:controller_1} and \ref{sec:controller_2} provide two controller designs based on a linear actuator and motor voltage respectively. Simulation results show time-series response of switching between attractors. Conclusions and application potentials are discussed in Section~\ref{sec:conclusion}.

\section{Energy Harvester Design}
\label{sec:system_design}

\subsection{Mechanical System}

\begin{figure}[ht]
	\centering
	\begin{minipage}{0.49\textwidth}
		\centering
		\includegraphics[width=\textwidth]{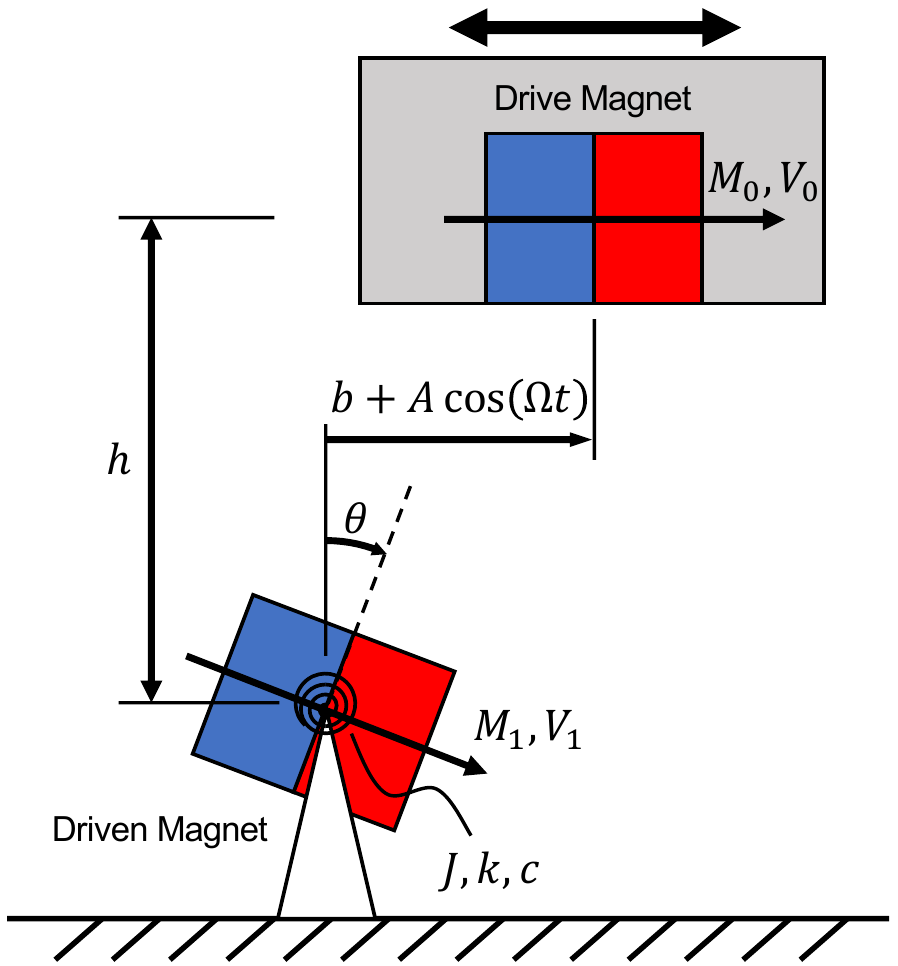}
		\caption{Schematic diagram of the non-contact translational-to-rotational magnetic transmission system.}
		\label{fig:schematic_mech}
	\end{minipage}
	\hfill
	\begin{minipage}{0.49\textwidth}
		\centering
		\includegraphics[width=0.9\textwidth]{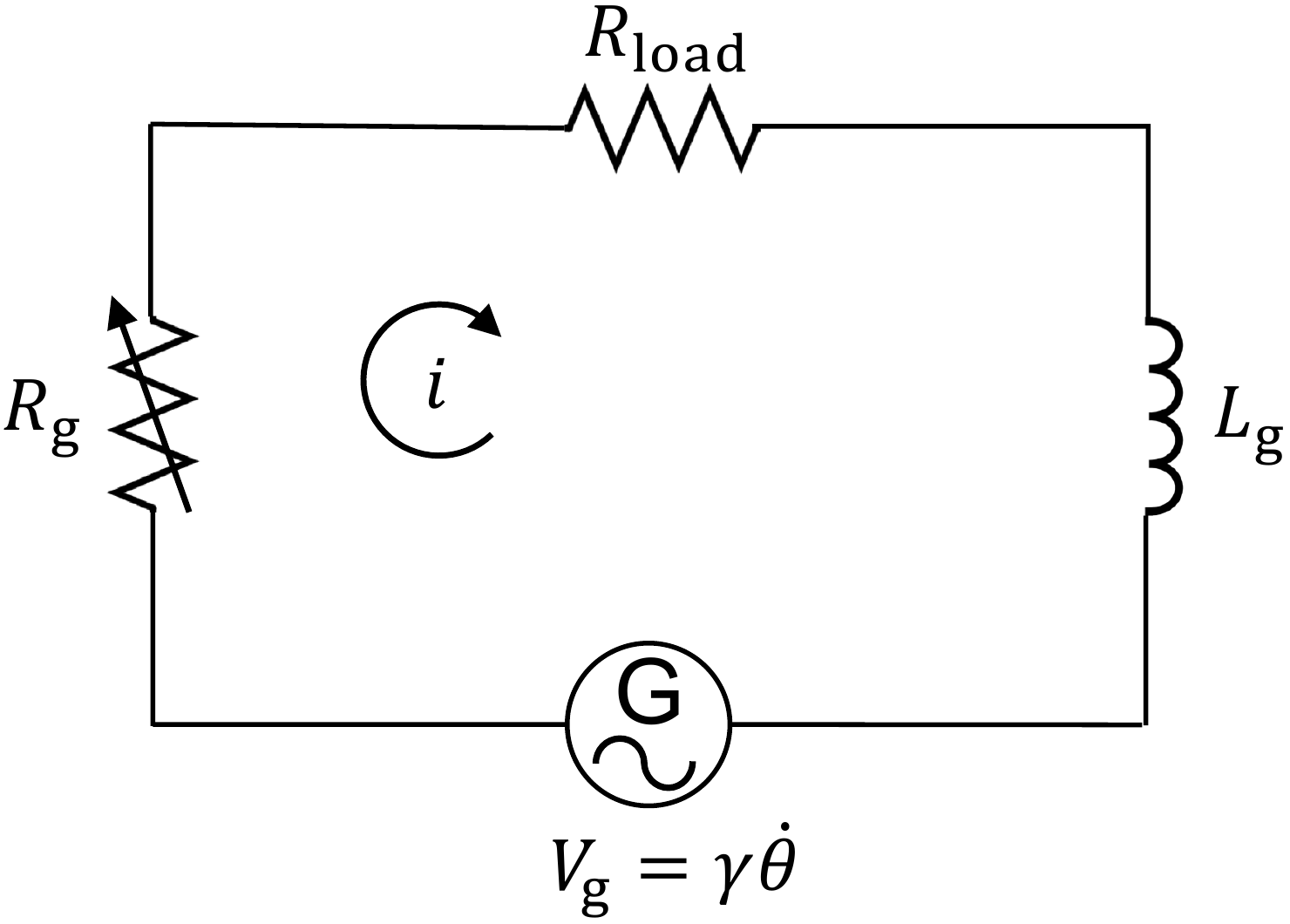}
		\caption{Schematic of the coupled electrical circuit in the generator. $R_g$ and $L_g$ are the resistance and the inductance inside the generator respectively. $V_g$ is the voltage induced by the rotary motion from the mechanical system.}
		\label{fig:circuit}
	\end{minipage}
\end{figure}

The energy harvester is based on a non-contact translational-to-rotational magnetic transmission in Ref.~\cite{wang2019nonlinear, wang2020dynamics}. As shown in Fig.~\ref{fig:schematic_mech}, the drive magnet is fixed to a vibration source and has a prescribed translational harmonic motion. The driven magnet is pinned and limited to pure rotation about its center of mass. The drive magnet applies a non-contact magnetic force to the driven magnet to achieve translational-to-rotational conversion. In addition to the magnetic force, the driven magnet is also subject to a linear restoring torque from mechanical springs and a damping torque which is assumed to be viscous and proportional to the driven magnet's angular velocity. 

As derived in Ref.~\cite{wang2019nonlinear}, the equation of motion can be written as a linear mass-spring-damper system driven by magnetic torque:
\begin{equation}
\begin{split}
J \ddot{\theta} + c \dot{\theta} + k \left( \theta - \theta_0 \right) = \qquad\qquad\qquad\qquad\qquad\qquad\qquad\\
\alpha \left[ \frac{\sin \theta}{\left[ \left( b + A \cos(\Omega t) \right) ^ 2 + h ^ 2 \right] ^ {3/2}} - \frac{3 \left[ b + A \cos(\Omega t) \right] \left[ h \cos \theta + \left( b + A \cos(\Omega t) \right)\sin \theta \right]}{\left[ \left( b + A \cos(\Omega t) \right) ^ 2 + h ^ 2 \right] ^ {5/2}} \right],
\end{split}
\label{eq:eom_mechanical_orig}
\end{equation}
where in the left-hand side, $J$, $c$, $k$ and $\theta_0$ are the driven magnet's moment of inertia, torsional spring coefficient, torsional damping coefficient, and offset bias angle of spring respectively. The right-hand side is the expression of magnetic torque, which is dependent on the angle of the driven magnet $\theta$, the vertical distance between two magnets $h$, and the horizontal distance between two magnets $b + A\cos(\Omega t)$. A constant $\alpha = \mu_0 M_0 V_0 M_1 V_1 / {4\pi}$ describes the magnetic properties of magnets, where $\mu_0$ is the permeability of free space, $M_0$ and $V_0$ are the drive magnet's magnetization and volume respectively, and $M_1$ and $V_1$ are the driven magnet's magnetization and volume respectively.

\subsection{Electro-Mechanical Coupling}

The mechanical system in Fig.~\ref{fig:schematic_mech} and Eq.~\eqref{eq:eom_mechanical_orig} transmits the vibrational source's translational motion to the driven magnet's rotational motion. This rotation is then conveyed to an electro-magnetic transducer, such as a generator, to produce electric power. For simplicity, as shown in Fig.~\ref{fig:circuit}, the voltage induced by the electro-magnetic transducer $V_g$ is assumed to be linearly dependent on the driven magnet's angular velocity $\dot{\theta}$, and the transducer connects to a simple resistor load $R_\text{load}$. As derived in Ref.~\cite{mann2012uncertainty}, the governing equation for the energy harvester is comprised of 1)~the mechanical system's governing equation Eq.~\eqref{eq:eom_mechanical_orig} where a coupling term $\gamma i$ was introduced, and 2)~an additional equation for the electrical circuit:
\begin{equation}
\begin{split}
J \ddot{\theta} + c \dot{\theta} + k \left( \theta - \theta_0 \right) - \boxed{\gamma i} = \qquad\qquad\qquad\qquad\qquad\qquad\qquad\\
\alpha \left[ \frac{\sin \theta}{\left[ \left( b + A \cos(\Omega t) \right) ^ 2 + h ^ 2 \right] ^ {3/2}} - \frac{3 \left[ b + A \cos(\Omega t) \right] \left[ h \cos \theta + \left( b + A \cos(\Omega t) \right)\sin \theta \right]}{\left[ \left( b + A \cos(\Omega t) \right) ^ 2 + h ^ 2 \right] ^ {5/2}} \right],
\end{split}
\label{eq:eom_mechanical}
\end{equation}
%
%
\begin{equation}
L_\text{g} \dot{i} + \left( R_\text{g} + R_\text{load} \right) i + \boxed{\gamma \dot{\theta}} = 0
\label{eq:circuit_simple}
\end{equation}
where $R_g$ and $L_g$ are the resistance and the inductance inside the generator respectively. $\gamma$ is the electro-mechanical coupling term and $i$ is the current induced by the rotary motion from the mechanical system. The resistor load $R_\text{load}$ was used to evaluate the power output. In order to ensure the consistency of the energy harvester throughout this paper, its parameters were measured experimentally and then used for the following numerical and analytical investigations. The value of parameters can be found in Tab.~\ref{tab:params}.
%





\begin{table}[]
	\centering
	\caption{Parameters of the energy harvester.}
	\label{tab:params}
	\begin{tabular}{lll}
		\hline
		Parameter                  & Symbol  & Value \\
		\hline
		\multicolumn{3}{c}{Driven Magnet}  \\ 
		\hline                                
		Driven magnet's moment of inertia & $J$ & $1.11 \times 10^{-6}$ $\text{kg} \cdot \text{m}^2$      \\
		Spring torsion coefficient & $k$   & $5.48 \times 10^{-3}$ N$\cdot$m/rad    \\
		Damping torsion coefficient& $c$   & $3.02 \times 10^{-6}$ N$\cdot$m$\cdot$s/rad    \\ 
		Natural frequency without damping & $\omega_{\text{n}}$ & 70.25 rad/s \\
		Offset bias angle & $\theta_0$ & 0 rad \\
		\hline
		\multicolumn{3}{c}{Excitation}      \\ 
		\hline
		Excitation amplitude & $A$ & 3 mm \\
		Excitation frequency & $\Omega$ & 50.24 rad/s (8 Hz)\\
		Excitation horizontal bias & $b$ & 0 mm \\
		Vertical distance between magnets & $h$ & 34 mm \\
		\hline
		\multicolumn{3}{c}{Magnet Properties}      \\ 
		\hline
		Permeability of free space & $\mu_0$ & $4\pi \times 10 ^ {-7}$ H/m \\
		Residual flux density      & $B_r$   & 1.32 Tesla          \\
		Magnetization              & $M_0,\, M_1 = B_r / \mu_0$ & $1.05 \times 10 ^ 6$ A/m    \\
		Magnet radius              & $r_{\text{mag}}$  & 6.35 mm (1/4 in)  \\
		Magnet height              & $h_{\text{mag}}$  & 12.7 mm (1/2 in)   \\
		Magnet volume              & $V_0,\, V_1 = \pi r_{\text{mag}}^2  h_\text{mag}$   & 1608.8 $\text{mm}^3$ \\ 
		\hline
		\multicolumn{3}{c}{Circuit}      \\ 
		\hline
		Generator inductance & $L_\text{g}$ & 1 H \\
		Generator resistance & $R_\text{g}$ & 0.1 $\Omega$ \\
		Electric load & $R_\text{load}$ & 5 $\Omega$ \\
		Coupling coefficient & $\gamma$ & 0.06 \\
		\hline
	\end{tabular}
\end{table}

\section{Coexisting Attractors}
\label{sec:attractors}

\begin{figure}[ht]
	\centering
	\includegraphics[width=\linewidth]{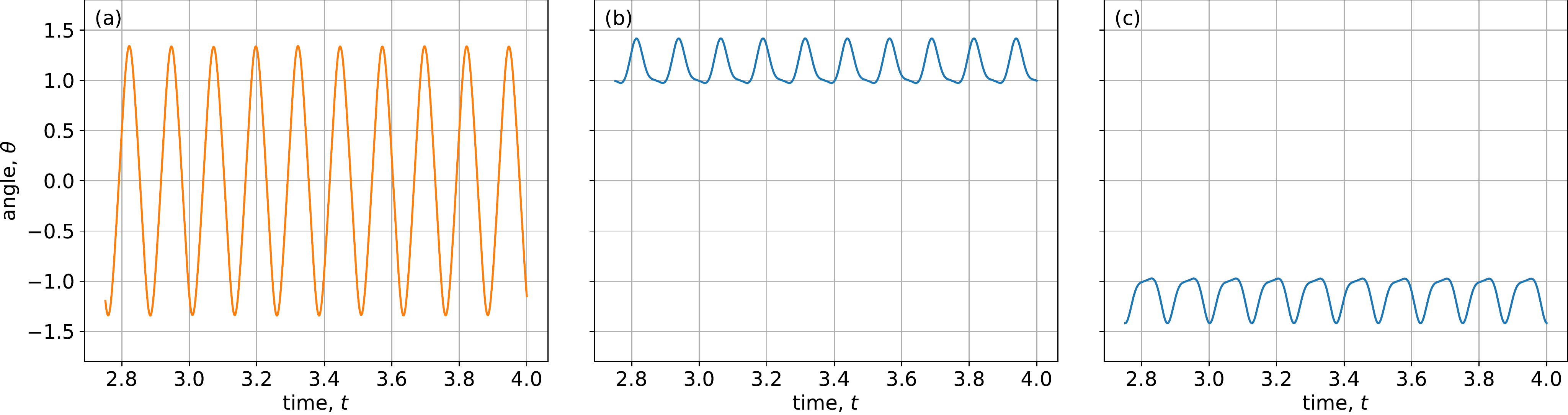}
	\caption{Coexisting steady-state oscillations of the driven magnet's angle $\theta$. (a) has a larger oscillation amplitude while (b) and (c) oscillate around symmetric offset angles with smaller amplitudes. They result from different initial conditions: $[\theta_0, \dot{\theta}_0, i_0]$ = (a) [-1.15, -38, 0.07], (b) [1.0, -1.4, 0.008], (c) [-1.0, 1.4, -0.008]. }
	\label{fig:attractor_tseries}
\end{figure}

The presence of coexisting attractors is a challenge for nonlinear energy harvesting. Integrating the governing equations Eq.~\eqref{eq:eom_mechanical} and \eqref{eq:circuit_simple} with respect to time results in steady-state solutions after initial transient dissipates. These solutions are stable thus they are also considered "attractors" in dynamical systems. As shown in Fig.~\ref{fig:attractor_tseries}, different initial conditions result in three coexisting steady-state oscillations: one with larger amplitude and two symmetric oscillations with smaller amplitudes. Their response frequencies are all equal to the excitation frequency (the excitation is the motion of vibrational source, $b+A\cos{\Omega t}$).

\begin{figure}[ht]
	\centering
	\includegraphics[width=0.9\linewidth]{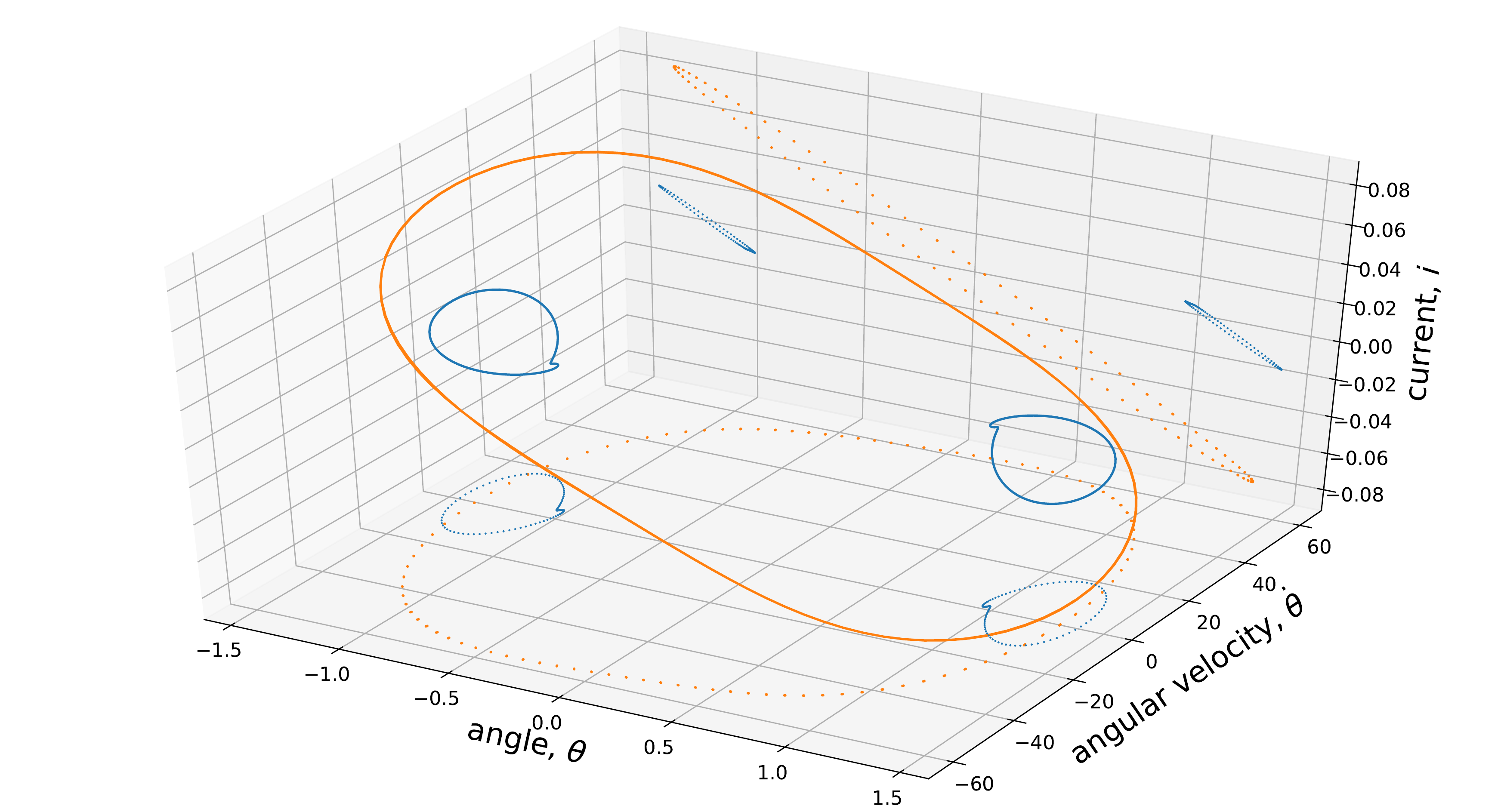}
	\caption{Phase portrait of stead-state oscillations of the energy harvester. The orange cycle with a larger oscillation amplitude represents high-power output (\textbf{HP} attractor) while the two symmetric blue cycles with smaller oscillation amplitudes represent low-power output (\textbf{LP} attractor). The dotted lines are the projections of attractors on 2D planes (top view and side view).}
	\label{fig:phase_3d}
\end{figure}

The energy harvester has three state variables: the driven magnet's angle $\theta$, angular velocity $\dot{\theta}$ and the induced current $i$. Thus the phase portraits of steady-state oscillations are closed cycles in 3-dimensional phase domain (see Fig.~\ref{fig:phase_3d}). The phase portraits include two symmetric smaller cycles and one larger cycle. Given that their response frequencies are equal, the cycle with the larger current amplitude corresponds to high-power output (considered \textbf{HP} attractor), while the two cycles with the smaller current amplitudes correspond to low-power output (considered \textbf{LP} attractor). 

\begin{figure}[ht]
	\centering
	\includegraphics[width=0.8\textwidth]{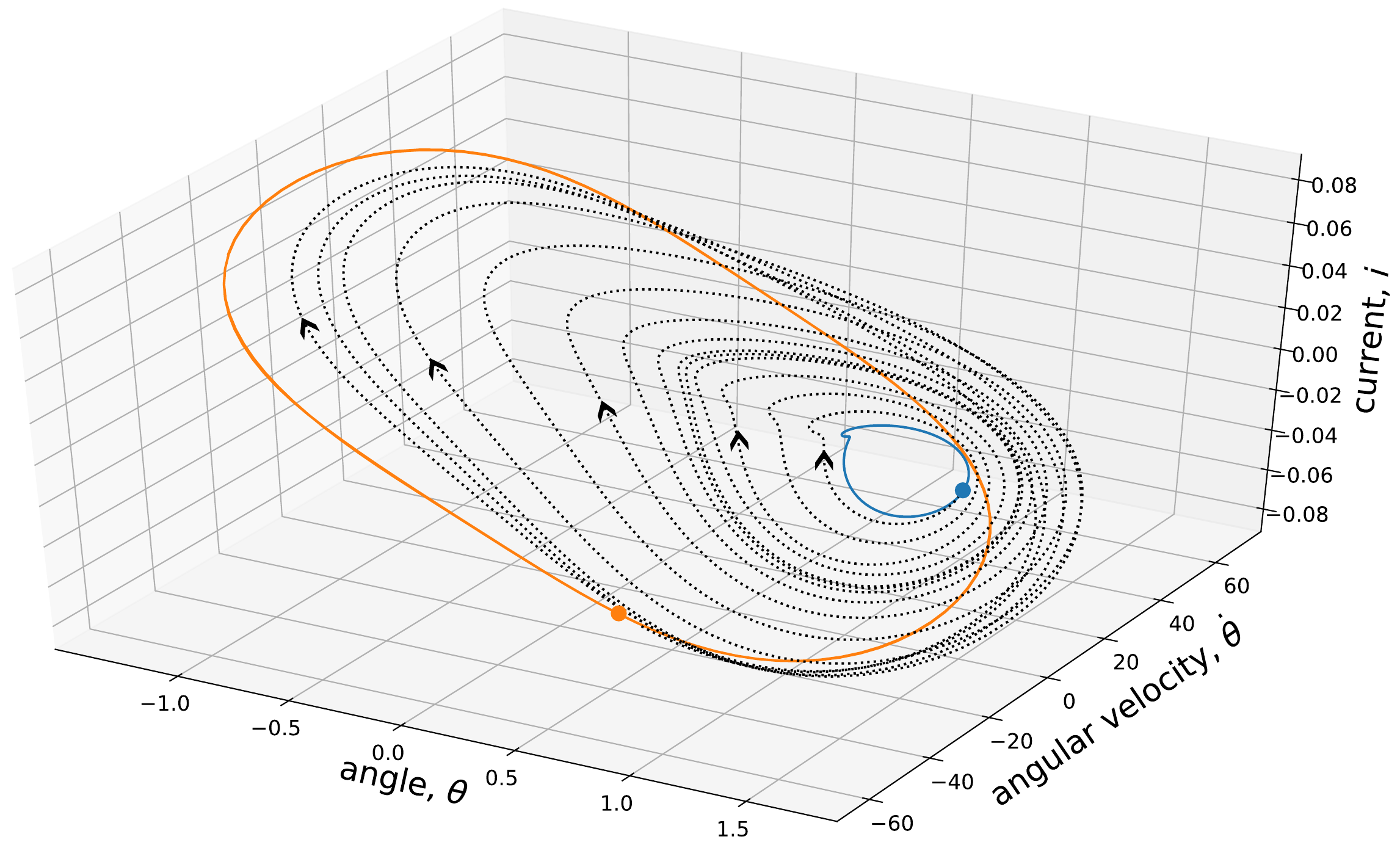}
	\caption{Attractor switching form LP to HP attractor by using the proposed control method.}
	\label{fig:phase_transient}
\end{figure}

As for an energy harvester, running on the HP attractor is generally more desirable, but sometimes LP attractor could also be needed for safety reasons or physical restrictions. The following sections will introduce a control method to switch between coexisting attractors of the energy harvester (see~Fig.\ref{fig:phase_transient}).


\section{Reinforcement Learning}
\label{sec:rl_framework}

\subsection{Framework}
Reinforcement Learning (RL), which can be viewed as an approach to optimal control of nonlinear systems, learns action-decisions while optimizing the long-term consequences of actions. In the RL framework, an \textit{agent} gains experience by making \textit{observations}, taking \textit{actions} and receiving \textit{rewards} from an \textit{environment}, and then learns a \textit{policy} from past experience to achieve goals (usually maximized cumulative \textit{reward}). This section implements RL algorithms for the energy harvester to switch between its low-power (LP) and high-power (HP) attractors using minimized energy consumed on the controller.

\textbf{Environment} is represented by the energy harvester's governing equation. As shown in Fig.~\ref{fig:attractor_tseries} and \ref{fig:phase_3d}, for certain ranges of the parameters, the energy harvester will always eventually settle into one of the stable attractors. Our objective is to apply control to the system to make it switch between LP and HP attractors using constrained actuation. To provide actuation for attractor switch, the velocity of the linear actuator is controlled by $a(s)$, which depends on the energy harvester's states $s$. 

\textbf{Action}. Aligned with the practical consideration that an actuation is commonly constrained, the action term can be written as $a(s) := F \pi_\eta(s)$, where $F$ is the action bound which denotes the maximum absolute value of the actuation, and $\pi_\eta(s)$ is the control policy. $\pi_\eta(s)$ is designed to be a function parameterized by $\eta$, which has an input of the energy harvester's states $s$, and an output of an actuation scaling value between $-1$ and $1$. Our objective is achieved by finding qualified parameters $\eta$ that cause the desired attractor to be reached.

\textbf{State \& Observation}. The energy harvester is driven by a time-varying excitation $A\cos{\Omega t}$; thus, the state should at least consist of time $t$, angle $\theta$, angular velocity $\dot{\theta}$, and induced current $i$. Given that $A\cos{\Omega t}$ is a sinusoidal function with a phase periodically increasing from $0$ to $2\pi$, time can be replaced by phase for a simpler state expression. The system's state can therefore be written as $s := \left[ \phi,\, \theta,\, \dot{\theta},\, i \right]$, where $\phi$ is equal to $\omega t \text{ modulo } 2 \pi$. Additional state variable(s) might be needed according to the definition of action $a$. Multiple definitions of the control input are introduced in the next section, where more details of state variables are discussed. For the sake of simplicity, we have assumed that no observation noise was introduced and the states were fully observable by the agent. 

\textbf{Reward}. A well-trained policy should use a minimized control effort (i.e. energy consumed on a controller) to reach the target attractor; thus the reward function, which is determined by state $s_t$ and action $a_t$, should inform 1)~whether the energy harvester reaches the target attractor, and 2)~the cost of the action taken. The environment estimates whether the target attractor will be reached by evaluating the next state $s_{t+1}$. A constant reward of $r_\text{end}$ is given only if the target attractor will be reached. Similar to the state in RL, the action cost, $r_\text{cost}$, differs in the definition of action $a$, thus the detail of calculating the cost is presented along with the specific controller design in the next section. A generic reward function for attractor selection can be written as:
\begin{equation}
r(s_t, a_t) = - r_\text{cost} +
\begin{cases}
r_\text{end}, & \text{if $s_{t+1}$ reaches the target attractor (or BoA)}\\
0,     & \text{otherwise}
\end{cases}
\end{equation}
For estimating whether the target attractor will be reached, one could observe whether $s_{t+1}$ is in the ``basin of attraction'' of the target attractor. Basins of attraction (BoA) are the sets of initial states leading to their corresponding attractors as time evolves. Once the target BoA is reached, the system will automatically settle into the desired stable trajectory without further actuation.

\subsection{Basin of Attraction (BoA) Prediction} 

Determining whether a state is in the target BoA is non-trivial. For an instantaneous state $s_t = [\phi_t,\, \theta_t,\, \dot{\theta}_t,\, i_t]$, we could integrate the governing equation with the initial condition $[\phi_0,\, \theta_0,\, \dot{\theta}_0,\, i_0]$. Integrating for a sufficiently long time should give a steady-state response, whose amplitude can be evaluated to determine the attractor where the system will eventually settle down. However, this prediction is needed for each time step of the RL learning process, and the integration time should be sufficiently long to obtain steady-state responses; thus this approach results in expensive computational workload and a slow learning process. As a result, a more efficient method was needed for determining which resting attractor corresponded to the system's state~\cite{wang2020modelfree}.

\begin{figure}[ht]
	\centering
	\includegraphics[width=\linewidth]{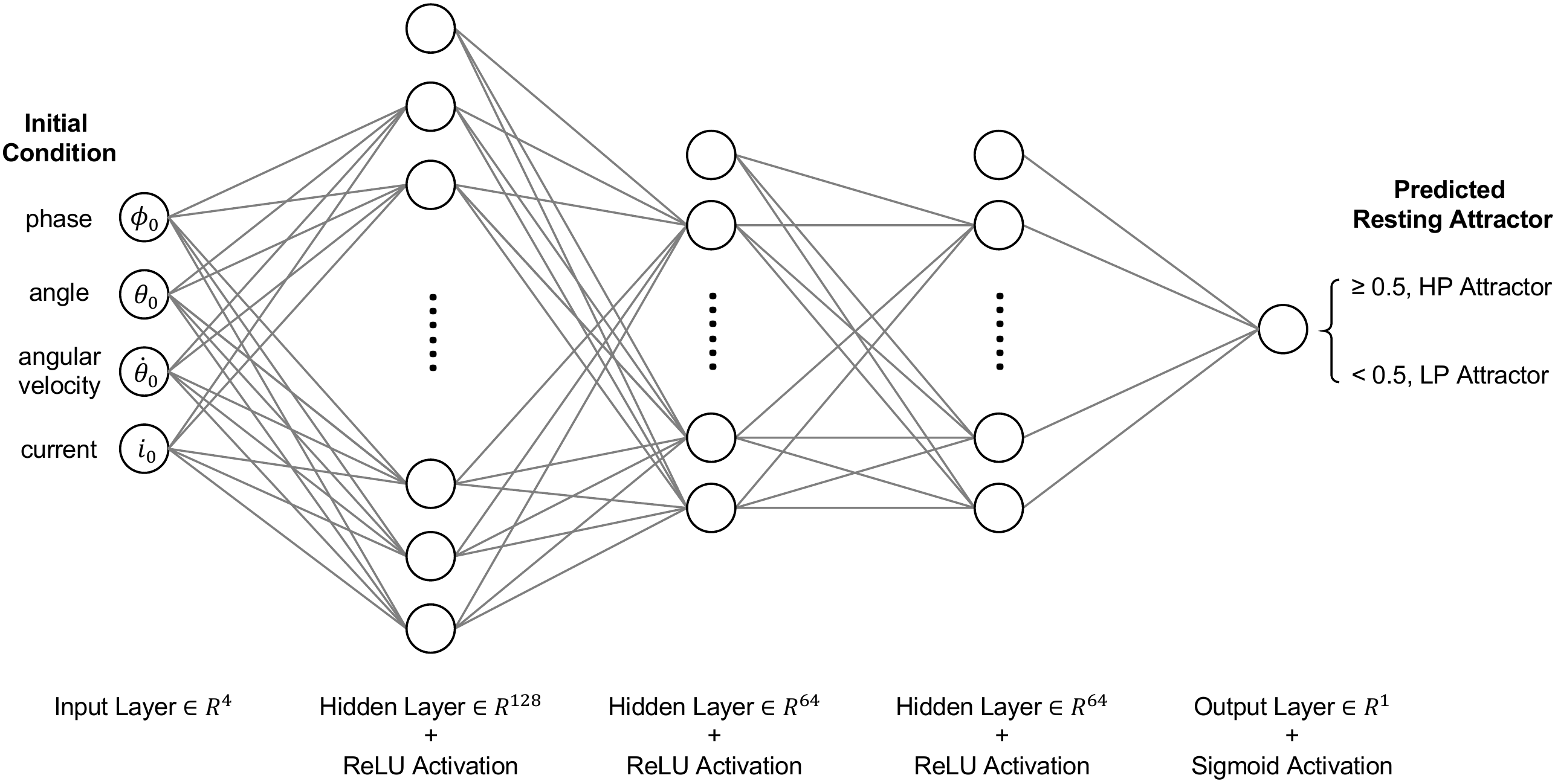}
	\caption{Neural network diagram for predicting the resting attractor (HP or LP) given an initial condition of the energy harvester $[\phi_0,\, \theta_0,\, \dot{\theta}_0,\, i_0]$. The network has four layers: 128 neurons (ReLU) -- 64 neurons (ReLU) -- 64 neurons (ReLU) -- 1 neuron (sigmoid). The ReLU (Rectified Linear Unit) activations introduce nonlinearity to this classifier. The sigmoid activation transforms its input into a value between 0.0 and 1.0, thus predicting the probability of HP attractor as an output. Therefore, when the output value is greater or equal than 0.5, the initial condition is predicted to lead the energy harvester to settle down on the HP attractor, otherwise the LP attractor.}
	\label{fig:nn_diagram}
\end{figure}

From the observation of coexisting attractors, there are two levels of power output for the energy harvester. Since the number of levels is finite, the attractor prediction can be considered a classification problem, where the input is the system's state and the output is the resting attractor where the system will settle down under no control. Given that the system's state has four dimension and the boundary of BoA tends to be nonlinear (or at least not guaranteed to be linear for the energy harvester), the classification algorithm using neural network was selected for its capability of handling high-dimensional nonlinear classification. As shown in Fig.~\ref{fig:nn_diagram}, a 4-layer neural network was trained (or called ``fitted'') to predict the resting attractor (HP or LP) given an initial condition $[\phi_0,\, \theta_0,\, \dot{\theta}_0,\, i_0]$. The training data was created by sampling states randomly on the domain of four state variables, and the corresponding resting attractor was determined by the method mentioned above: evaluating future responses with long-term integration of governing equation. Generally speaking, this method transfers the recurring cost of integration during the RL learning process to a one-time cost before the learning process begins. Collecting training data can be time consuming (running millions of simulation), but once the classifier is well-trained, the time for predicting the resting attractor can be negligibly small.  

\subsection{Algorithm}

\setcounter{magicrownumbers}{0}
\begin{table*}
	\caption{Algorithm: Deep Deterministic Policy Gradient (DDPG) for learning a control policy for attractor switching}
	\label{table:DDPG}
	\begin{tabular}{r|ll}
		\toprule
		\rownumber & Determine the governing equation: Eq.~\eqref{eq:eom_control_spr} for the actuator control, or Eq.~\eqref{eq:eom_control_vol} for the voltage control &\\
		\rownumber & Determine the reward function: Eq.~\eqref{eq:reward1} for the  actuator control, or Eq.~\eqref{eq:reward2} for the voltage control &\\
		\rownumber & Randomly initialize actor network ${\pi _\eta }(s)$ and critic network $Q_\psi(s, a)$ with weights $\eta$ and $\psi$ &\\
		\rownumber & Initialize target network $\pi'_{\eta'}(s)$ and $Q'_{\psi'}(s, a)$ with weights $\eta' \leftarrow \eta$, $\psi' \leftarrow \psi$ &\\
		\rownumber & Set the initial condition of the Duffing equation $s_0 = [x_0,\, \dot{x}_0,\, \phi_0]$ &\\
		\rownumber & Set discount factor $\gamma$, and soft update factor $\tau$ &\\
		\rownumber & Set time of Phase 1 and Phase 2, $T_1$ and $T_2$ &\\
		\rownumber & Initialize replay buffer $B$ &\\
		\rownumber & \textbf{for} episode = 1 : M \textbf{do} &\\
		\rownumber & \quad Initialize a random process $\mathcal{N}$ for action exploration &\\
		\rownumber & \quad Add noise to time of Phase 1, ${T_1}'=T_1+\text{random}(0, 2\pi / \omega)$ &\\
		\rownumber & \quad Integrate the governing equation for $t \in \left[0,\, {T_1}' \right]$ with $a(s)=0$ & Phase 1\\
		\rownumber & \quad\textbf{for} $t = {T_1}'$ : ${T_1}' + T_2$ \textbf{do} & Phase 2\\
		\rownumber & \quad\quad Observe current state, $s_t = \left[ x_t,\, \dot{x}_t,\, \phi_t \right]$ &\\
		\rownumber & \quad\quad Evaluate action $a_t(s_t) = F\pi_\eta(s_t) + \mathcal{N}_t$, according to the current policy and exploration noise &\\
		\rownumber & \quad\quad Step into the next state $s_{t+1}$, by integrating the governing equation for $\Delta t$ &\\
		\rownumber & \quad\quad Evaluate reward $r_t(s_t, a_t)$ from the reward function &\\
		\rownumber & \quad\quad Store transition $[s_t,\, a_t,\, r_t,\, s_{t+1}]$ in $B$ &\\
		\rownumber & \quad\quad Sample a random minibatch of $N$ transitions $[s_i,\, a_i,\, r_i,\, s_{i+1}]$ from $B$ &\\
		\rownumber & \quad\quad Set $y_i = r_i + \gamma Q'_{\psi'}(s_{i+1}, F\pi'_{\eta'}(s_{i+1}))$ &\\
		\rownumber & \quad\quad \makecell[l]{Update the critic network by minimizing the loss: \\ \quad\quad\quad\quad $L=\frac{1}{N}\sum\limits_i (y_i - Q_\psi(s_i, a_i))^2$} &\\
		\rownumber & \quad\quad \makecell[l]{Update the actor network using the sampled policy gradient: \\ \quad\quad\quad\quad $\nabla_\eta J \approx \frac{1}{N}\sum\limits_i \nabla_a Q_\psi(s, a) |_{s=s_i, a=F\pi_\eta(s_i)} \nabla_\eta \pi_\eta(s)|_{s=s_i}$   } &\\
		\rownumber & \quad\quad \makecell[l]{Update the target networks: \\ \quad\quad\quad\quad $\psi' \leftarrow \tau \psi + (1-\tau) \psi',\, \eta' \leftarrow \tau \eta + (1-\tau) \eta'$} &\\
		\rownumber & \quad\quad \textbf{if} the target attractor's basin is reached in $s_{t+1}$: \textbf{break}&\\
		\rownumber & \quad \textbf{end for} &\\
		\rownumber & \textbf{end for} & \\
		\bottomrule
	\end{tabular}
\end{table*}

Amongst various RL algorithms, Deep Deterministic Policy Gradient (DDPG) was selected for its capability of operating over continuous state and action spaces~\cite{lillicrap2015continuous}. The goal of DDPG is to learn a policy which maximizes the expected return $J = \mathbb{E}_{r_i, s_i, a_i}[R_{t=1}]$, where the return from a state is defined as the sum of discounted future rewards $R_t = \sum\nolimits_{i = t}^T \gamma^{i-t} r(s_i, a_i)$ with a discount factor $\gamma \in [0,\,1]$.

An action-value function, also known as a ``critic'' or ``Q-value'' in DDPG, is used to describe the expected return after taking an action $a_t$ in state $s_t$:
\begin{equation}
Q(s_t, a_t) = \mathbb{E}_{r_{i \geqslant t}, s_{i>t}, a_{i>t}}[R_t | s_t,a_t].
\end{equation}
DDPG uses a neural network parameterized by $\psi$ as a function appropriator of the critic $Q(s, a)$, and updates this critic by minimizing the loss function of the difference between the ``true'' Q-value $Q(s_t, a_t)$ and the ``estimated'' Q-value $y_t$:
\begin{equation}
L( \psi ) = \mathbb{E}_{s_t, a_t, r_t} \left[ \left( Q( s_t, a_t | \psi) - y_t \right) ^ 2 \right],
\end{equation} 
where
\begin{equation}
y_t = r(s_t, a_t) + \gamma Q( s_{t+1}, \pi( s_{t+1} )_t | \psi).
\label{eq:y_t}
\end{equation}
Apart from the ``critic'', DDPG also maintains an ``actor'' function to map states to a specific action, which is essentially our policy function $\pi(s)$. DDPG uses another neural network parameterized by $\eta$ as a function approximator of the actor $\pi(s)$, and updates this actor using the gradient of the expected return $J$ with respect to the actor parameters $\eta$:
\begin{equation}
\nabla_\eta J \approx \mathbb{E}_{s_t}\left[ \nabla_a Q_\psi(s, a) |_{s=s_t, a=\pi(s_t)} \nabla_\eta \pi_\eta (s) |_{s=s_t} \right].
\label{dJ}
\end{equation}	

In order to enhance the stability of learning, DDPG uses a ``replay buffer'' and separate ``target networks'' for calculating the estimated Q-value $y_t$ in Eq.~\eqref{eq:y_t}. The replay buffer stores transitions $[s_t,\, a_t,\, r_t,\, s_{t+1}]$ from experienced trajectories. The actor $\pi(s)$ and critic $Q(s, a)$ are updated by randomly sampling a minibatch from the buffer, allowing the RL algorithm to benefit from stably learning across uncorrelated transitions. The target networks are copies of actor and critic networks, $\pi'_{\eta'}(s)$ and $Q'_{\psi'}(s, a)$ respectively, that are used for calculating the estimated Q-value $y_t$. The parameters of these target networks are updated by slowly tracking the learned networks $\pi_{\eta}(s)$ and $Q_{\psi}(s, a)$:
\begin{equation}
\begin{split}
&\psi' \leftarrow \tau \psi + (1-\tau) \psi', \\
&\eta' \leftarrow \tau \eta + (1-\tau) \eta',
\end{split}
\end{equation}
where $0 < \tau \ll 1$. This soft update constrains the estimated Q-value $y_t$ to change slowly, thus greatly enhancing the stability of learning. 

In order to implement DDPG to attractor switching for the energy harvester, several terms needs be defined:
\begin{enumerate}
	\item[] \textit{Phase 1}: the phase where the system is free of control, i.e., $a(s)=0$. The system is given a random initial condition at the beginning of Phase 1, waits for dissipation of the transient, and settles down on the initial attractor at the end of Phase 1.
	\item[] \textit{Phase 2}: the phase following Phase 1; the system is under control. Phase 2 starts with the system running in the initial attractor, and ends with either reaching the target BoA or exceeding the time limit. 
	\item[] \textit{Trajectory}: the system's time series trajectories for Phase 2: $[s_t,\, a_t,\, r_t,\, s_{t+1}]$. The data of trajectories are stored in a replay buffer.
	\item[] \textit{Replay Buffer}: an implementation of experience replay~\cite{lin1992self}, which randomly selects previously experienced samples to update a control policy. Experience replay stabilizes the RL learning process and reduces the amount of experience required to learn~\cite{mnih2015human}.
\end{enumerate}
The entire process for learning a control policy can be summarized as iterative episodes. In each episode, the system runs through Phase 1 and Phase 2 in turn and the control policy is improved using the data of trajectories stored in the replay buffer. Deep deterministic policy gradient (DDPG) is a RL algorithm for updating the control policy.

The detailed learning process using DDPG can be found in Tab.~\ref{table:DDPG}. In line 11, the time of Phase 1 is perturbed by adding a random value between $0$ and $2\pi/\omega$ (a excitation period). This noise provides diversity of the system's states at the beginning of Phase 2, which enhances generality and helps prevent over-fitting of the control policy network $\pi_\eta(s)$. Both the actor network $\pi_\eta(s)$ and the critic network $Q_\psi(s, a)$ have two hidden layers, each of which has 128 units and an activation function of ReLU \cite{glorot2011deep}. For the actor network, the final output layer is a tanh layer to bound the actions. For the critic network, the input layer consists of only the state $s$, while the action $a$ is included in the 2nd hidden layer. Adam optimizer \cite{kingma2014adam} was used to learn the neural network parameters with a learning rate of $\tau_\eta = 10^{-4}$ and $\tau_\psi = 10^{-3}$ for the actor and critic respectively. For the update of the critic network we used a discount factor of $\gamma = 0.9$. For the soft update of the target network $\pi'_{\eta'}(s)$ and $Q'_{\psi'}(s, a)$ by Polyak Averaging, we used $\tau = 0.1$. For the system's settling down in Phase 1 we used $T_1 = 2$, and for constraining the time length of control we used $T_2 = 4$. The replay buffer had a size of $10^6$. In each episode, the minibatch of transitions sampled from the replay buffer had a size of $N=64$.

\begin{figure}[ht]
	\centering
	\begin{minipage}{0.35\textwidth}
		\centering
		\includegraphics[width=\textwidth]{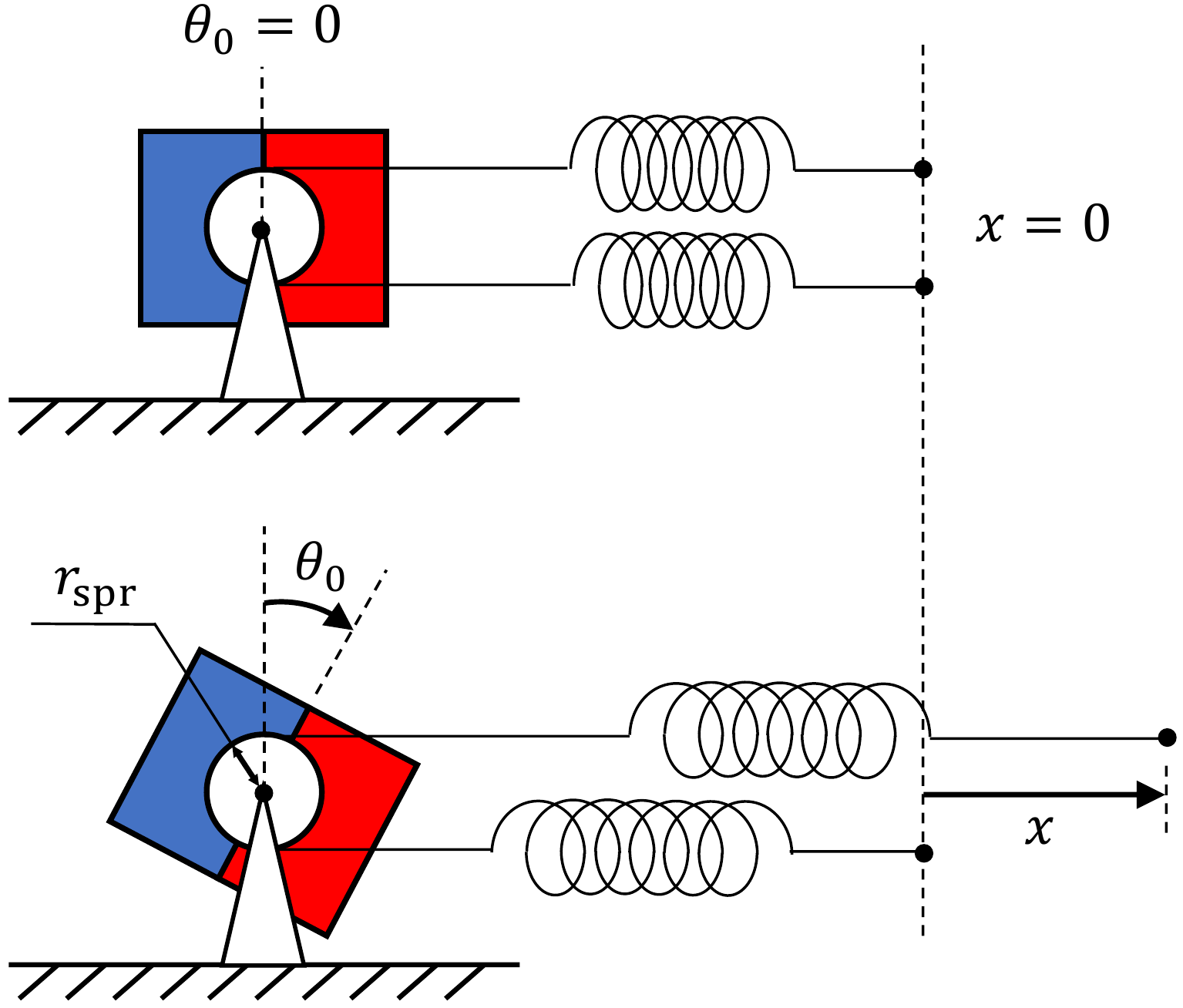}
		\caption{Static equilibria of the offset bias angle $\theta_0$. It is proportional to the distance between the two springs' ends: $x = 2 r_\text{spr} \theta_0$.}
		\label{fig:spring}
	\end{minipage}
	\hfill
	\begin{minipage}{0.6\textwidth}
		\centering
		\includegraphics[width=\textwidth]{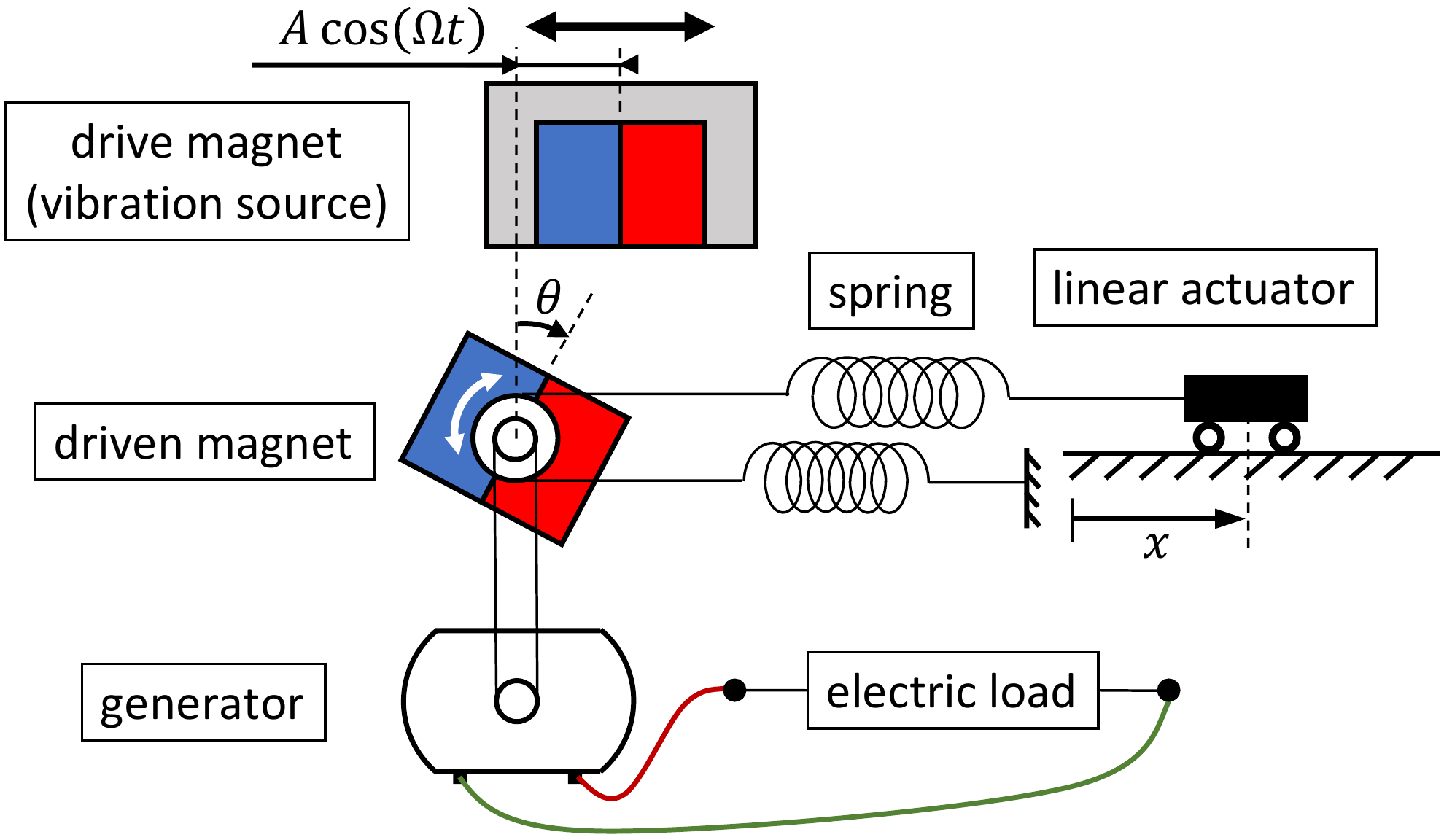}
		\caption{Schematic of control option I, where an external torque on the driven magnet is provided by moving spring position with a linear actuator. The driven magnet transmits rotational motions to a generator to produce electric power. Figure~\ref{fig:circuit} shows the electro-mechanical coupling circuit.}
		\label{fig:schematic}
	\end{minipage}
\end{figure}

\section{Control Option I: Linear Actuator for Moving Spring Position}
\label{sec:controller_1}

\subsection{Controller Design}

Given that the driven magnet exhibits rotational oscillations while the energy harvester is operating, an reasonable control input is an external torque on the driven magnet. Reforming the governing equation Eq.~\eqref{eq:eom_mechanical} leads to:
\begin{equation}
\begin{split}
I \ddot{\theta} + c \dot{\theta} + k \theta - \gamma i &= \tau_\text{mgt} + \boxed{k \theta_0}, \\
L_\text{g} \dot{i} + \left( R_\text{g} + R_\text{load} \right) i + \gamma \dot{\theta} &= 0, \\
\end{split}
\label{eq:eom_control_torque}
\end{equation}
where $\tau_\text{mgt}$ is the magnetic torque, i.e., the right hand side of Eq.~\eqref{eq:eom_mechanical}. The term $k\theta_0$ was intentionally moved to the right hand side, thus can be considered an external torque that is controllable. It forms a torque determined by the offset bias angle $\theta_0$, which is the static equilibrium angle when the system is only forced by springs. As shown in the Fig.~\ref{fig:spring}, the restoring torque in the system is provided by two linear springs wrapped around the circular plate beneath the driven magnet. The offset bias angle $\theta_0$ can be altered by moving springs, and has a linear relationship with the distance between ends of springs $x$:
\begin{equation}
x = 2 r_\text{spr} \theta_0,
\end{equation}
where $r_\text{spr}$ is the radius of the circular plate wrapped by springs. In order to manipulate the distance $x$, as shown in Fig.~\ref{fig:schematic}, the end of one spring is fixed and that of the other spring is attached to a linear actuator. Instead of directly controlling the actuator's position $x$, its velocity $\dot{x}$ was chosen to be the control input for a more practical control scenario. The system's governing equation becomes:
\begin{equation}
\begin{split}
I \ddot{\theta} + c \dot{\theta} + k \theta - \gamma i &= \tau_\text{mgt} + \frac{k}{2 r_\text{spr}} x, \\
L_\text{g} \dot{i} + \left( R_\text{g} + R_\text{load} \right) i + \gamma \dot{\theta} &= 0, \\
\dot{x} &= a,
\end{split}
\label{eq:eom_control_spr}
\end{equation}
where $a$ is the control input -- the actuator's velocity. Instead of the conventional notation of control input $u$ in control theory, research in reinforcement learning (RL) uses $a$ to represent ``actions''. This RL-style notation are used throughout the remainder of this paper.

\subsection{Reward Function}
As discussed in Sec.~\ref{sec:rl_framework}, the reward function should give 1)~a constant positive reward $r_\text{end}$ when energy harvester reaches a terminal successfully, and 2)~penalties $r_\text{cost}$ (negative rewards) for the cost of actions taken. Evaluating the action cost is tricky, which ideally should be the electric energy consumed on the actuator. However, a realistic actuator is a complex electro-mechanical coupling system, thus making it difficult to calculate its energy consumption especially for a dynamic load. In addition, it's hard to find a generic model to represent all various actuators. 

For simplicity, the action cost is defined as the work done by the actuator, that is, ``the force pulling on the spring end'' times ``the distance the spring end moves''. According to Hooke's law, the force can be expressed as $k_\text{spr} (x_t - r_\text{spr} \theta_t + x_0)$, where $k_\text{spr}$ is the stiffness of the linear spring, $x_t - r_\text{spr}$ is the distance stretched, and $x_0$ is a positive constant to ensure springs are alway stretched. The distance the spring end moves can be written as $a_t \Delta t$, where $a_t$, the control input, is the actuator's instantaneous velocity, and $\Delta t$ is the time step size for control. The work done by the actuator is therefore $k_\text{spr} (x_t - r_\text{spr} \theta_t + x_0) a_t \Delta t$. However, this calculation of the action cost is counter-intuitive when $a_t < 0$. The negative $a_t$ leads to a negative work done by the actuator, that is, the actuator is powered by its load. To avoid this paradox, the negative value of $a_t$ was cut off and the reward function becomes:
\begin{equation}    
r(s_t, a_t) = - k_\text{spr} (x_t - r_\text{spr} \theta_t + x_0) \text{Max} ({a}_t,\, 0) \Delta t +
\begin{cases}
r_\text{end}, & \text{if $s_{t+1}$ reaches target BoA}\\
0,     & \text{otherwise}
\end{cases}
\label{eq:reward1}
\end{equation}
It's worth reiterating that the action cost calculated in this reward function is not the energy consumed on a realistic actuator, but it can still be considered the least energy needed for an action in each time step. 

\begin{figure}[ht]
	\centering
	\includegraphics[width=0.88\linewidth]{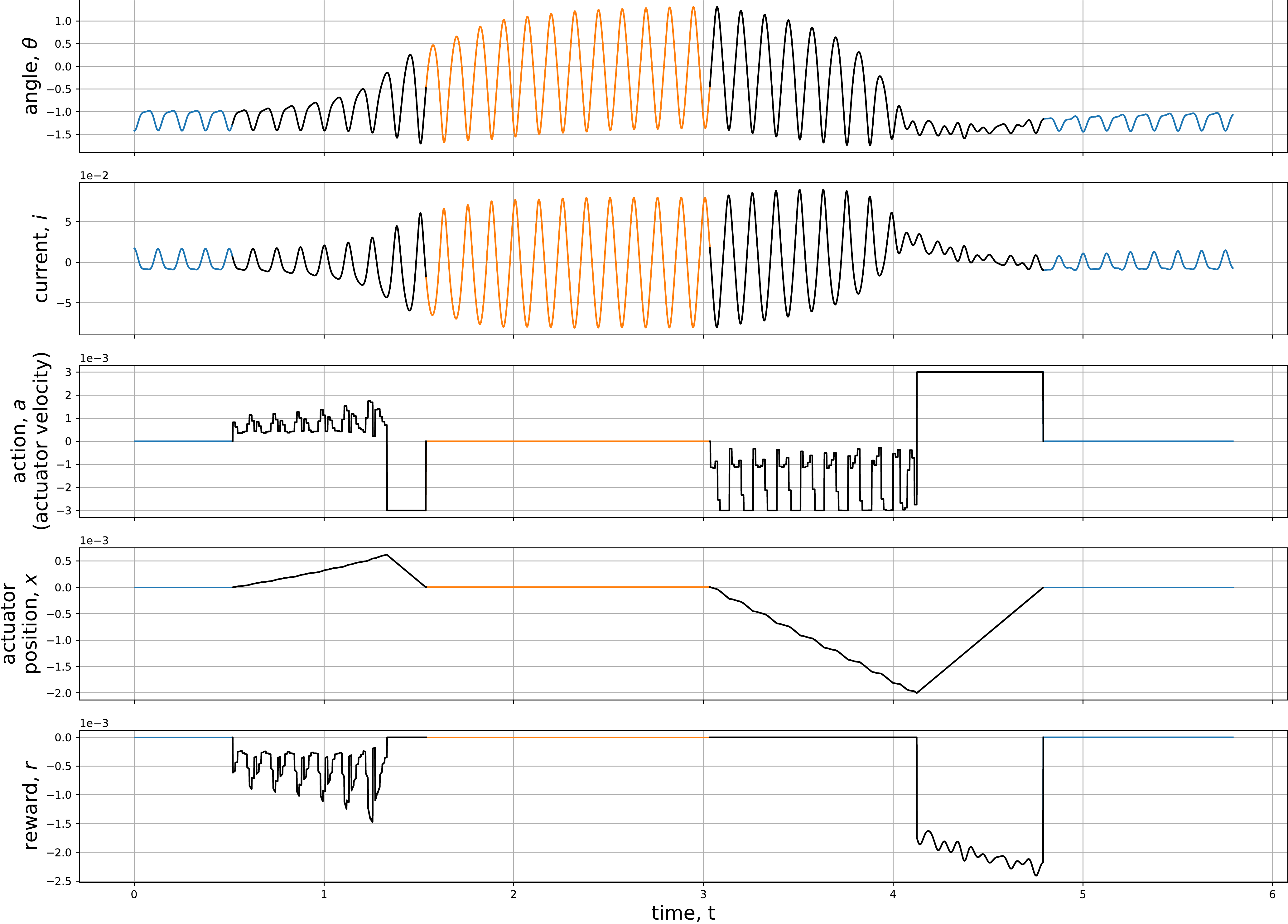}
	\caption{Time series of the attractor switching using the control option I (spring control) + DDPG. The energy harvester governed by Eq.~\eqref{eq:eom_control_spr} was first switched from LP to HP attractor and then switched back from HP to LP attractor. The blue, orange and black lines represent the system running on the LP attractor, running on the HP attractor, being under control, respectively.}
	\label{fig:spr_ddpg}
\end{figure}
\begin{figure}[ht]
	\centering
	\includegraphics[width=0.88\linewidth]{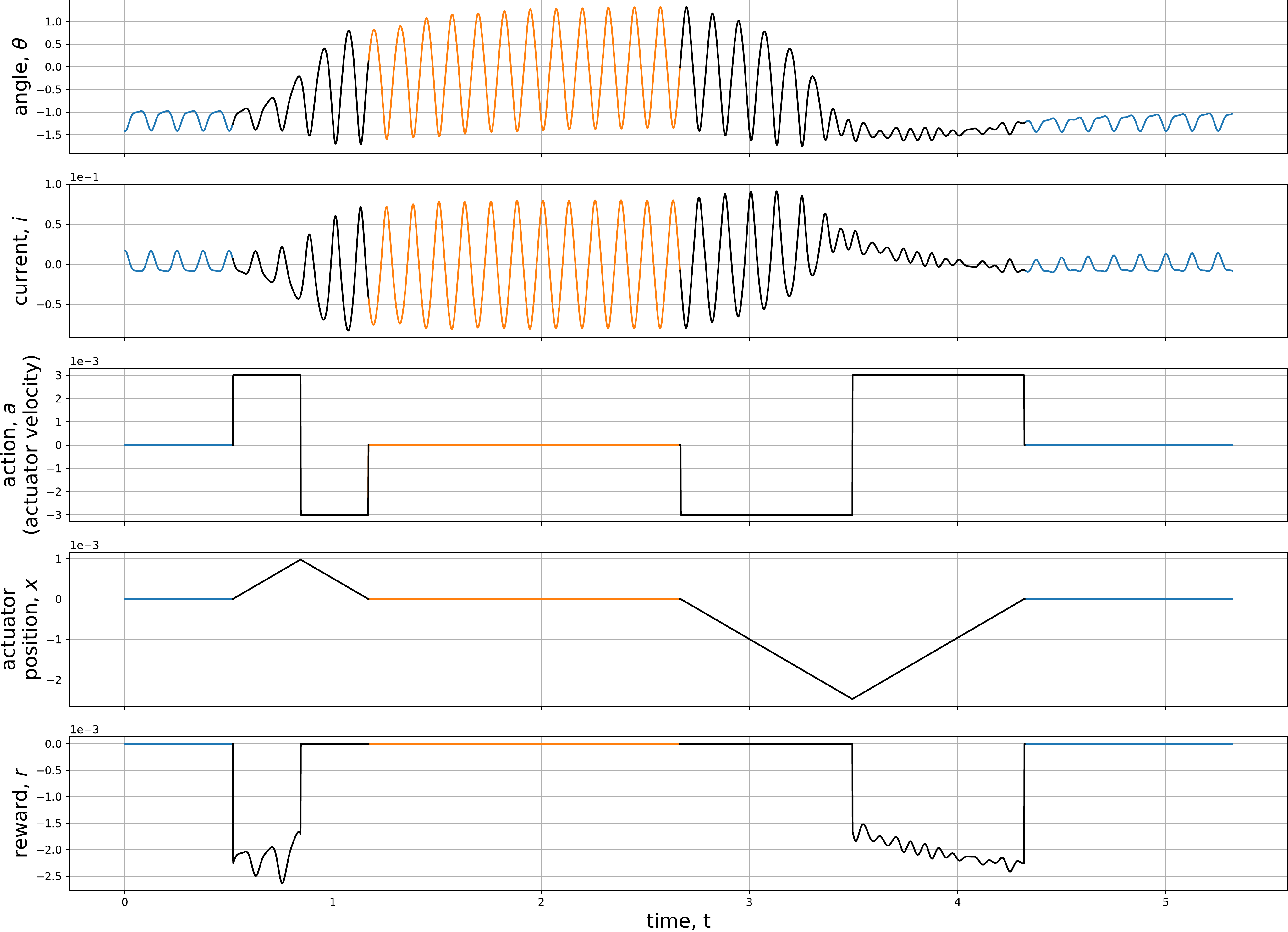}
	\caption{Time series of the attractor switching using the control option I (spring control) + quasi-bang-bang control. The energy harvester governed by Eq.~\eqref{eq:eom_control_spr} was first switched from LP to HP attractor and then switched back from HP to LP attractor. The blue, orange and black lines represent the system running on the LP attractor, running on the HP attractor, being under control, respectively.}
	\label{fig:spr_bang}
\end{figure}

\subsection{Results}

Figure~\ref{fig:spr_ddpg} shows the time series of successful attractor switching using the control policy learned by RL. As observed in Sec.~\ref{sec:attractors}, the attractor with a small amplitude of steady-state response (blue lines) is named ``LP'' (short for low-power) while that with a large amplitude (orange lines) is named ``HP'' (short for how-power). It's worth noting that each policy only controls a one-way trip of attractor switching. One control policy is needed for transitioning from the LP to HP, and another control policy is needed for the reverse direction. Therefore, two independent sets of parameter $\eta$ were learned for the actor functions $a(s)=F\pi_\eta(s)$. In addition, $F$ is the action bound, which represents the maximum absolute velocity allowed for the actuator. It was set 0.003 m/s for both directions of attractor switching.

Figure~\ref{fig:spr_ddpg} includes five phases:
\begin{enumerate}
	\item controller OFF, operating on the LP attractor (blue lines).
	\item controller ON, switching from LP to HP (black lines).
	\item controller OFF, waiting for the dissipation of transient process, then operating on the HP attractor (orange lines).
	\item controller ON, switching from HP to LP (black lines).
	\item controller OFF, waiting for the dissipation of transient process, then operating on the LP attractor (blue lines).
\end{enumerate}
Each time when the controller is ON, one can observe that the action plot consists of two stages: first jagged curve (control policy learned by RL) and then a step function followed (manually-designed control policy). This results from the fact that a successful attractor switching is not only the system's reaching the target BoA, but also returning the actuator back to its original position, i.e.~$x = 0$. If it were otherwise ($x \neq 0$), the attractors themselves will be changed for a non-zero offset bias angle $\theta_0$ (recall that the coexisting attractors observed in Sec.~\ref{sec:attractors} is based on $\theta_0 = 0$ in Tab.~\ref{tab:params}). Therefore, after the control policy learned by RL drives the energy harvester to the target BoA, additional control policy is needed to reset the actuator. This also explains the trend of actuation position $x$, where the curve goes away from the origin at first and then returns back. 

The reward plot in Fig.~\ref{fig:spr_ddpg} demonstrates the work done by the actuator. The area above the curve represents the amount of work, which shows the HP-to-LP switching ($1.35 \cross 10^{-3}$ J) needs more energy than the LP-to-HP switching ($4.39 \cross 10^{-4}$ J). Furthermore, as defined in Eq.~\eqref{eq:reward1}, work is done only when the actuator is moving right ($a_t > 0$). This explains that the cost only exists in first half of LP-to-HP switching and second half of HP-to-LP switching. 

Although RL proves to realize attractor switching using a linear actuator, the jagged motion of action leads to rapid changes of the actuator's velocity, which might not work for low-sensitivity actuators. Another control policy with more smooth action was proposed by imitating the trend of the actuator's position in RL control. For simplicity, the action can only be switched between three states: OFF, ON with maximum positive velocity, and ON with maximum negative velocity. This controller design is similar to a bang-bang controller with binary states (ON \& OFF), thus named as ``quasi-bang-bang control''. As shown in Fig.~\ref{fig:spr_ddpg}, the trend of the actuator position is approximately linear. Given that the linear function derivative is a constant, the similar trend of actuator position in RL control can be imitated by implementing constant actuator velocities in quasi-bang-bang control. 

Figure~\ref{fig:spr_ddpg} shows the time series of successful attractor switching using the quasi-bang-bang control. When the controller is turned on, the actuator starts moving with the maximum velocity. Once the energy harvester reaches the target BoA, the actuator turns to the opposite direction and keeps moving with the maximum velocity until returning to the zero position. LP-to-HP switching requires positive actions first and then negative actions, while HP-to-LP switching has the reverse order. This brute-force control method provides a more smooth control than RL, but the lack of optimization over control results in larger energy consumption. By accumulating the rewards in Fig.~\ref{fig:spr_ddpg}, the HP-to-LP switching needs $1.69 \cross 10^{-3}$ J while the LP-to-HP switching needs $7.62 \cross 10^{-4}$ J.

In summary, this section proposes two control methods based on moving the spring position using a linear actuator. The control policy learned by RL optimizes energy consumption yet shows jagged motions of the actuator, while the quasi-bang-bang control, which was designed by imitating and simplifying the RL control policy, provides a more smooth control input yet lacks in optimization of energy consumption. However, even the RL control gives only an ``incomplete'' optimization. The requirement of returning the actuator to its original position divides the control process into two stages, and RL only optimizes the energy consumption in the first stage. Therefore, in the next section, another control method is proposed for optimizations over the whole control process.
%

%


\section{Control Option II: External Voltage on the Motor}
\label{sec:controller_2}

\begin{figure}[ht]
	\centering
	\begin{minipage}{0.55\textwidth}
		\centering
		\includegraphics[width=\textwidth]{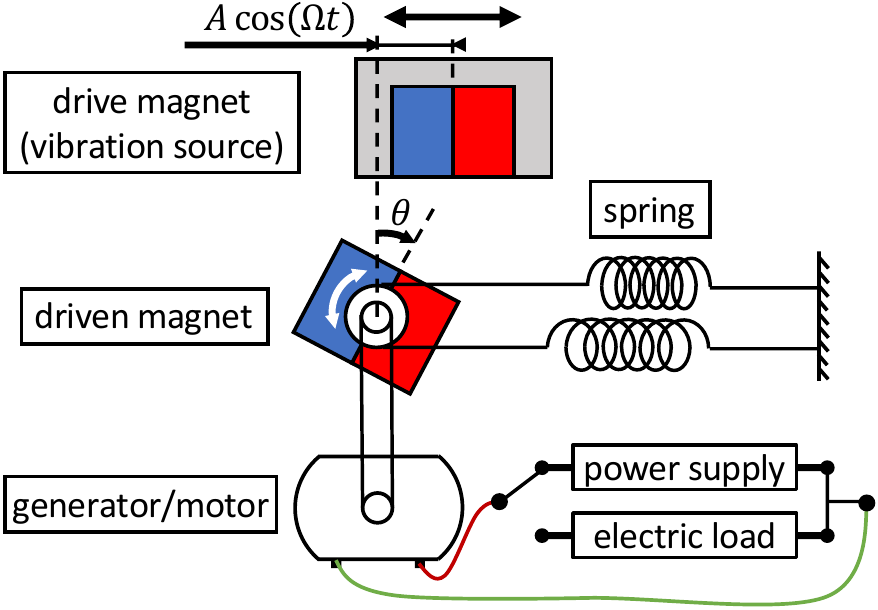}
		\caption{Schematic of control option II (voltage control).}
		\label{fig:schematic_voltage}
	\end{minipage}
	\hfill
	\begin{minipage}{0.4\textwidth}
		\centering
		\includegraphics[width=\textwidth]{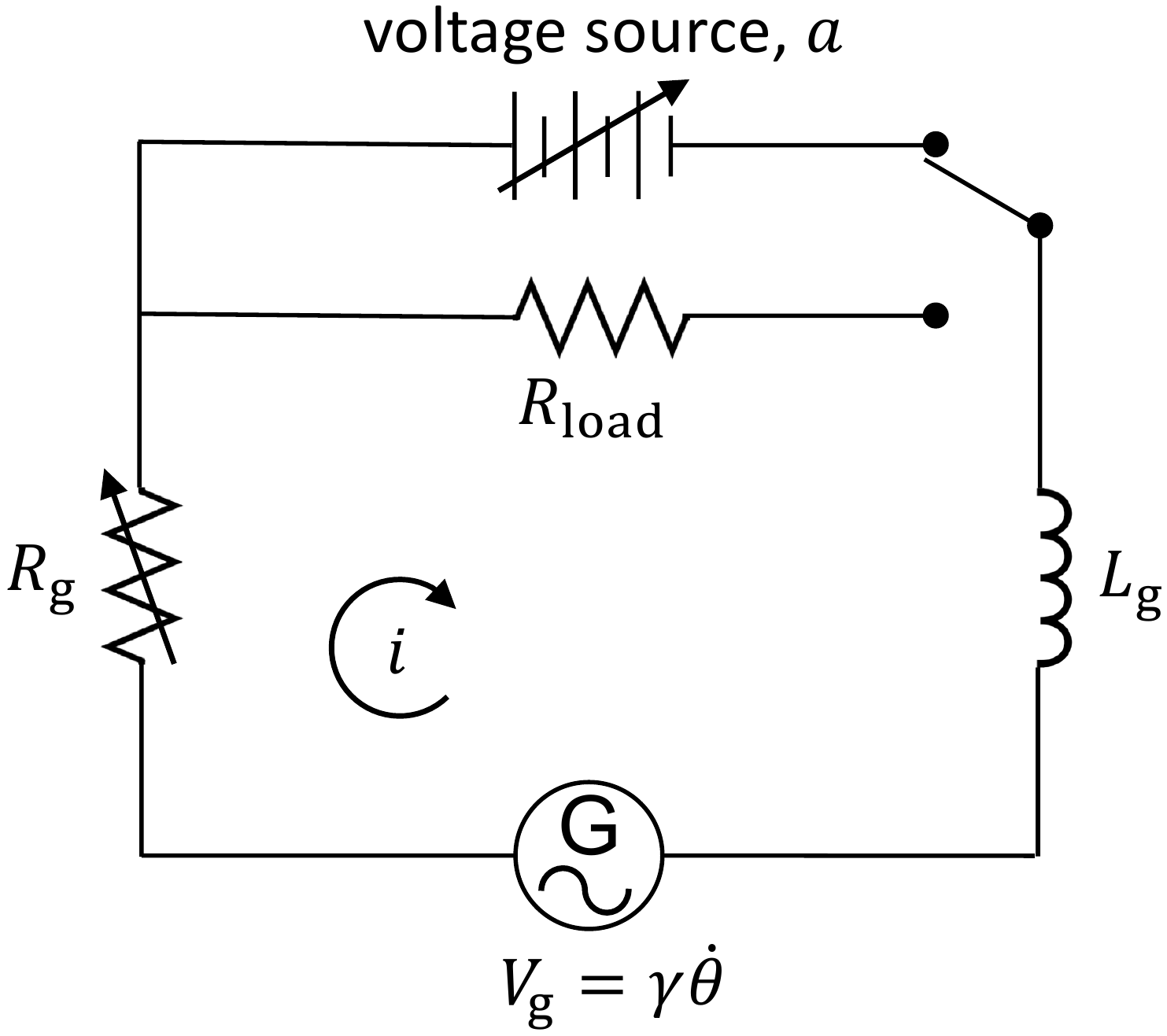}
		\caption{Circuit of control option II (voltage control).}
		\label{fig:circuit_control}
	\end{minipage}
\end{figure}

\subsection{Controller Design}

The energy harvester's attractors are steady-state oscillations determined by three state variables: the driven magnet's angle, angular velocity and induced current. Apart from exerting an external torque to control the angle and velocity using an actuator in Sec.~\ref{sec:controller_1}, the induced current could also be controlled by introducing an external power supply in the electro-magnetic circuit. As shown in Fig.~\ref{fig:schematic_voltage} and \ref{fig:circuit_control}, while the energy harvester is operating on an attractor, the generator is driven by the rotating magnet and powering an electric load. When the energy harvester is going to switch to another attractor, the generator is detached from the electric load and connected to a power supply, thus becoming a ``motor'' to reversely drive the rotating magnet. The circuit will be switched back to connect the electric load once the target attractor is reached. 

The system's governing equation becomes:
\begin{equation}
\centering
\begin{split}
I \ddot{\theta} + c \dot{\theta} + k \theta - \gamma i &= \tau_\text{mgt}, \\
L_\text{g} \dot{i} +  R_\text{g} i + \gamma \dot{\theta} &= 
\begin{cases}
a, & \text{if controller is ON}\\
- R_\text{load} i, & \text{if controller is OFF}
\end{cases}
\end{split}
\label{eq:eom_control_vol}
\end{equation}
where $\tau_\text{mgt}$ is the magnetic torque, i.e., the right hand side of Eq.~\eqref{eq:eom_mechanical}, and $a$ is the control input of voltage. The circuit is switched by using a relay as shown in Fig.~\ref{fig:circuit_control}. 

\subsection{Reward Function}
As discussed in Sec.~\ref{sec:rl_framework}, the reward function should give 1)~a constant positive reward $r_\text{end}$ when energy harvester reaches a terminal successfully, and 2)~penalties $r_\text{cost}$ (negative rewards) for the cost of actions taken. Evaluating the action cost is tricky, which ideally should be the input power of the power supply. However, the complex circuit for a realistic power supply makes it a difficult calculation, especially for a time-varying voltage output. In addition, there is no generic mathematical model to represent various power supplies, which may come from electric power grids, energy storage devices, generators/alternator, etc. 

For simplicity, the action cost is defined as the power supply's output power, that is, ``the voltage of control input $a_t$'' times ``the current flowing through the generator/motor $i_t$''. The action cost for each time step is therefore $a_t i_t \Delta t$. However, this calculation of the action cost is unrealistic for negative power $a_t i_t < 0$, which means the power supply is being charged by the energy harvester. To avoid this paradox, it's assumed that the power supply has an internal protection mechanism to cut off the negative power. The reward function becomes:
\begin{equation}
r(s_t, a_t) = - \text{Max} ({a}_t i_t ,\, 0) \Delta t +
\begin{cases}
r_\text{end}, & \text{if $s_{t+1}$ reaches target BoA}\\
0,     & \text{otherwise}
\end{cases}
\label{eq:reward2}
\end{equation}
Again, it's worth reiterating that the action cost calculated in this reward function is not the energy consumed on a realistic power supply, but it can still be considered the least energy needed for attractor switching using a voltage control. 

\subsection{Results}

\begin{figure}[ht]
	\centering
	\includegraphics[width=0.8\textwidth]{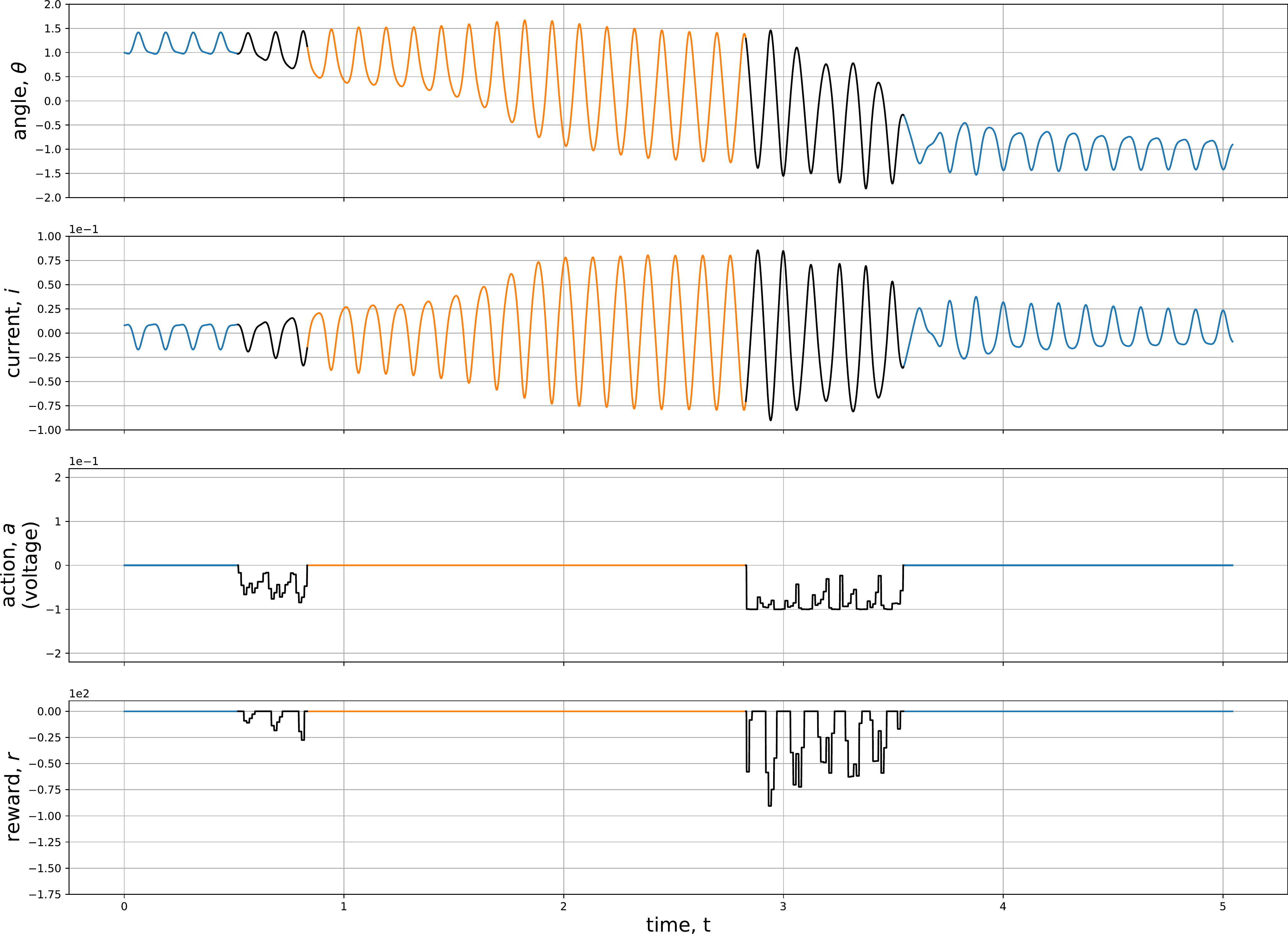}
	\caption{Time series of the attractor switching using the control option II (motor voltage) + voltage bound $F = 0.1$. The energy harvester governed by Eq.~\eqref{eq:eom_control_vol} was first switched from LP to HP attractor and then switched back from HP to LP attractor. The blue, orange and black lines represent the system running on the LP attractor, running on the HP attractor, being under control, respectively.}
	\label{fig:vol_bound_10e-2}
\end{figure}

\begin{figure}[h]
	\centering
	\includegraphics[width=0.8\textwidth]{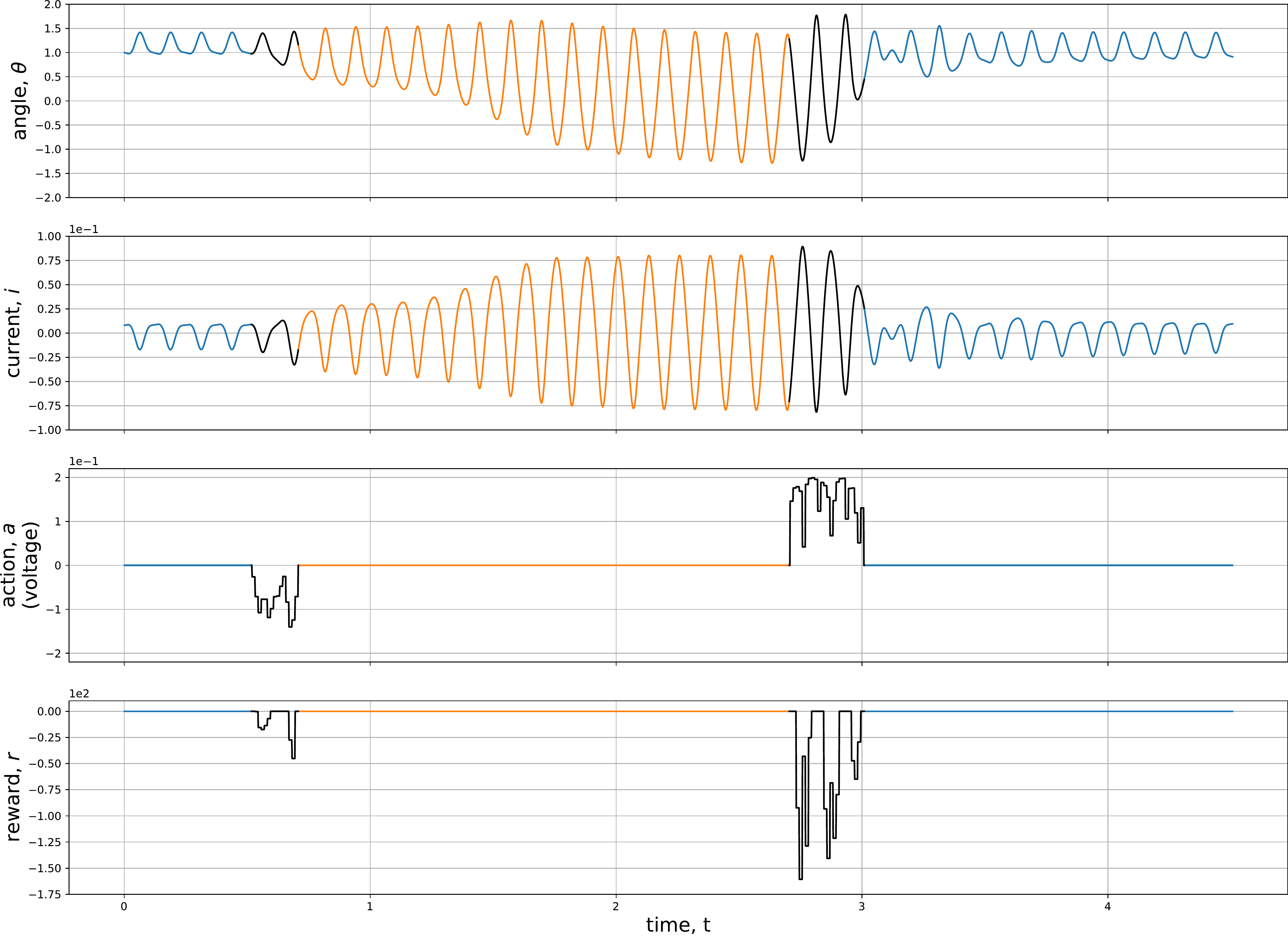}
	\caption{Time series of the attractor switching using the control option II (motor voltage) + voltage bound $F = 0.2$. The energy harvester governed by Eq.~\eqref{eq:eom_control_vol} was first switched from LP to HP attractor and then switched back from HP to LP attractor. The blue, orange and black lines represent the system running on the LP attractor, running on the HP attractor, being under control, respectively.}
	\label{fig:vol_bound_20e-2}
\end{figure}

For the attractor switching case that used voltage control, constraints were constructed with different action bounds, i.e., maximum voltage $F$. Recall that the control term in Eq.~\eqref{eq:eom_control_vol} can be written as $a(s) = F\pi_\theta(s)$, which is bounded between $-F$ and $F$. Figure~\ref{fig:vol_bound_10e-2} and \ref{fig:vol_bound_20e-2} show the time series of successful attractor switching using the control policy learned by RL, and they have the action bounds of $F = 0.1$ V and $F = 0.2$ V respectively.


Each time-series trajectory has five stages, including LP-to-HP and HP-to-LP switching:
\begin{enumerate}
	\item controller OFF, operating on the LP attractor (blue lines).
	\item controller ON, switching from LP to HP (black lines).
	\item controller OFF, waiting for the dissipation of transient process, then operating on the HP attractor (orange lines).
	\item controller ON, switching from HP to LP (black lines).
	\item controller OFF, waiting for the dissipation of transient process, then operating on the LP attractor (blue lines).
\end{enumerate}
Comparing between two action bounds, one can observe that the smaller bound results in a longer time length of control. It can be qualitatively explained by the energy threshold for jumping from one attractor to another. The energy provided by the power supply should be accumulated beyond the energy threshold to push the system away from one attractor. A smaller action bound therefore leads to longer time for the energy accumulation. 

Another observation is from the duration of control process. Compared with the short time length for control (region of black lines), the attractor switching spends more time waiting for dissipation of the transient process (especially for the LP-to-HP switching in both Fig.~\ref{fig:vol_bound_10e-2} and \ref{fig:vol_bound_20e-2}), where the system is automatically approaching the target attractor under no control. This phenomenon shows a smart and efficient strategy the control policy has learned by RL: instead of driving the system precisely to the states of the target attractor, it just drives the system to the attractor's basin, where the system might be far away from the target attractor initially but will reach it without further control effort as time evolves.

\subsection{Comparison}

The reward plots in Fig.~\ref{fig:vol_bound_10e-2} and \ref{fig:vol_bound_20e-2}) demonstrate the power supply's output power. The area above the curve represents the total energy consumption for the attractor switching. Although the LP-to-HP switching obviously spends less energy than the HP-to-LP switching, the comparison between two action bounds is obscure. In order to compare the energy consumption quantitatively, their values were calculated by integrating rewards with respective to time. Furthermore, the performance of the energy harvester was evaluated by using the amount of energy harvested in each forcing period, i.e., $E = \int_T i^2 R_\text{load}$, where $T$ is the forcing period of vibrational energy source ($T = 2\pi/\Omega$). Given that the response frequencies of both LP and HP attractor is equal to the forcing frequency, this calculation is capable of representing the average level of energy harvesting. As a result, in each forcing period, the energy harvested in LP attractor is $E_\text{LP} = 5.294 \cross 10^{-5}$ J while the energy harvested in HP attractor is $E_\text{HP} = 1.861 \cross 10^{-3}$ J.

\begin{figure}[h]
	\centering
	\includegraphics[width=0.8\textwidth]{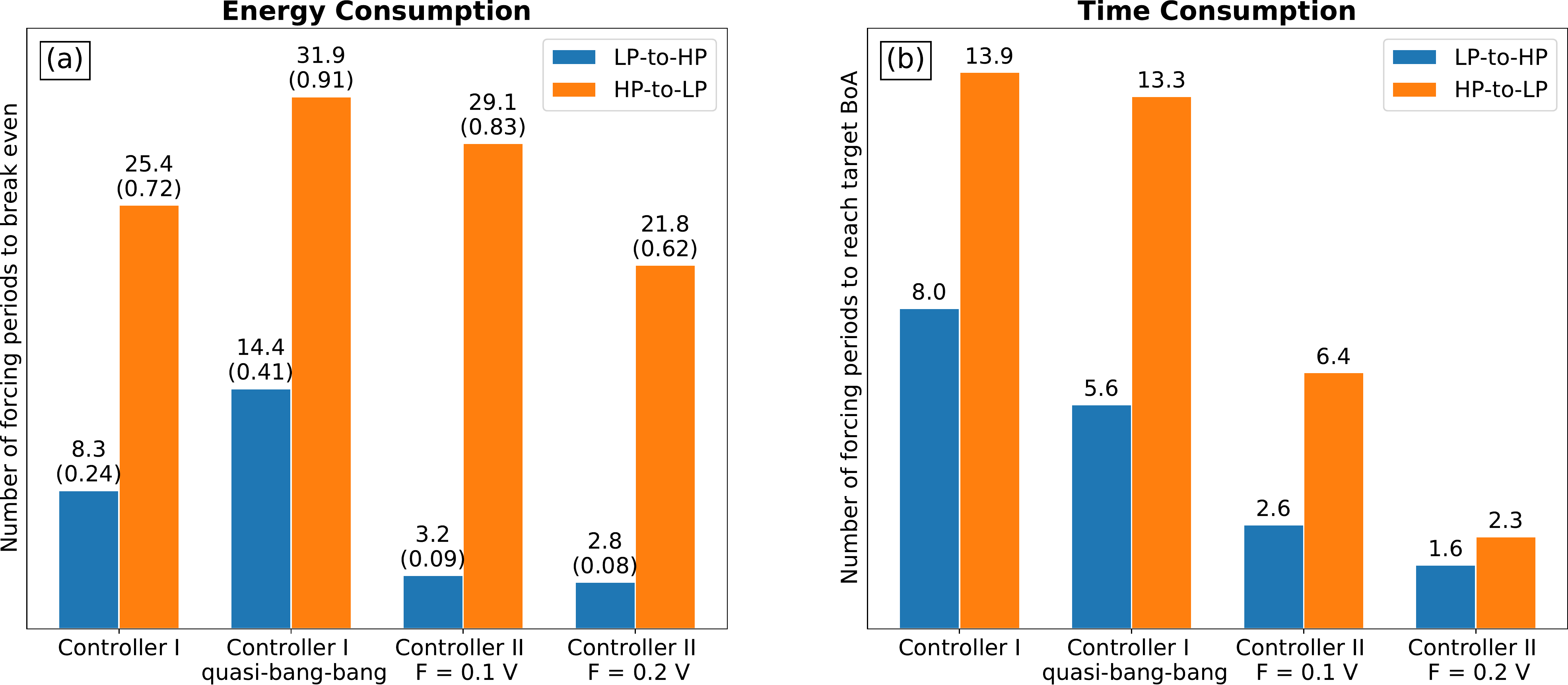}
	\caption{Energy and time consumption for attractor selection in energy harvesting. (a) The energy consumption is represented by the number of forcing periods to break even, which was calculated by dividing the total energy consumed for attractor switching by the energy harvested in each forcing period $E_\text{LP}$ (or $E_\text{HP}$). (b) The time consumption is represented by the number of forcing periods to reach target BoA, which was calculated by dividing the time length of control by the forcing period.}
	\label{fig:comparison}
\end{figure}

Figure~\ref{fig:comparison} compares the energy and time consumption for attractor selection under various control scenarios. In Fig.~\ref{fig:comparison}(a), the energy consumption is represented by the number of forcing periods to break even, which was calculated by dividing the total energy consumed for attractor switching by the energy harvested in each forcing period ($E_\text{LP}$ or $E_\text{HP}$). There are two numbers above each bar: the top one is the number of forcing periods needed to break even when running on the HP attractor while the bottom one in the parenthesis is for the system running on the LP attractor. In Fig.~\ref{fig:comparison}(b), the time consumption is represented by the number of forcing periods to reach target BoA, which was calculated by dividing the time length of control by the forcing period ($2\pi / \Omega$). In order to present a fair comparison, the energy and time consumption for each control scenario is the average of 50 times of attractor switching with random initial conditions. 


Several qualitative analysis can be performed from Fig.~\ref{fig:comparison}. In general, a large energy consumption corresponds to a longer control duration. For a certain controller, the HP-to-LP switching always consumes more energy than the LP-to-HP switching. The controller II (voltage control) is more energy efficient than the controller I (spring control) for LP-to-HP switching, while it has no obvious advantage in the HP-to-LP switching. For the controller I, the simplified quasi-bang-bang control consumes more energy than the original one learned by reinforcement learning, which agrees with our analysis in Sec.~\ref{sec:controller_1}. For the controller II with different action bounds (maximum absolute voltage), the smaller bound (i.e. more strict constraint) results in more energy consumption. 

\section{Conclusion}
\label{sec:conclusion}

This paper introduces a nonlinear energy harvester based on a translation-to-rotational magnetic transmission; the harvester's behavior contains coexisting attractors with different levels of electric power output. Two controller designs were investigated to switch the energy harvester's response from one attractor to another. The first controller applies external torques on the driven magnet by altering the position of the spring that provides restoring force. The spring position is controlled by a linear actuator. The second design controls the voltage on the generator in the energy harvester, which temporarily converts the generator into a motor to affect the system's dynamics. 

The control policies based on the two controllers were learned by using deep deterministic policy gradient (DDPG) -- a deep reinforcement learning (RL) method. A RL framework for attractor switching was presented, which defines the environment, action, state \& observation and reward function. In addition, a neural network classifier was used to rapidly predict the resting attractor based upon the system's instantaneous state. 

Both control methods obtained from RL successfully switch the energy harvester's response between coexisting attractors. For the control with an actuator, another quasi-bang-bang control was proposed by imitating and simplifying the RL-learned policy. This simplified method is less energy-efficient yet gives more smooth motion of the actuator, which may benefit a realistic actuator. For the voltage control, the RL-learned control policies prove to work under different action bounds (i.e. a power supply's maximum voltage). 

Future work needs to extend our investigations to experimental implementation. Especially for the energy consumption, its measurement on a real linear actuator (controller I) or power supply (controller II) is needed to evaluate the practical performance of an energy harvesting system. Other factors that have not been investigated in this paper (such as observation noise, communication delay, nonlinear damping, etc.) should also be considered in experiments. In order to realize attractor selection experimentally, a feasible solution is to first find a sub-optimal control policy in simulation, as performed in this paper, and then “transfer” the pre-trained policy to the experiments for further optimization. This process is called “transfer learning”, where the heavy-learning workload in experiments is shared with simulations. More studies implementing the attractor selection approach based on real-world experiments and transfer learning are certainly worthy topics for further investigations.



\bibliographystyle{unsrt}  
\bibliography{references.bib}  

\end{document}